\documentclass[aps, a4paper,superscriptaddress, nofootinbib, showpacs, twocolumn, 10pt]{revtex4}
\usepackage{amsmath,amssymb, bm}
\usepackage{dsfont}
\usepackage{epsfig}
\usepackage{graphicx}
\usepackage{slashed}
\usepackage{color}
\usepackage{subfigure} 
\newcommand{\be}{\begin{equation}}
\newcommand{\ee}{\end{equation}}

\newcommand{\beq}{\begin{equation}}
\newcommand{\eeq}{\end{equation}}
\newcommand{\bea}{\begin{eqnarray}}
\newcommand{\eea}{\end{eqnarray}}
\newcommand{\ba}{\begin{align}}
\newcommand{\ea}{\end{align}}
\newcommand{\bfig}{\begin{figure}}
\newcommand{\efig}{\end{figure}}

\newcommand{\wh}{\widehat}
\newcommand{\nn}{\nonumber}

\newcommand{\ah}{\hat a}

\begin{document}
~\vspace{1cm}
\title{ Renormalization-scheme variation of a QCD perturbation expansion with tamed large-order behavior}
\author{Irinel Caprini}
\affiliation{Horia Hulubei National Institute for Physics and Nuclear Engineering,
P.O.B. MG-6, 077125 Bucharest-Magurele, Romania}
 
\begin{abstract} The renormalization-scheme  and scale dependence of the truncated QCD perturbative  expansions is one of the main sources of theoretical error of the standard model predictions, especially at intermediate energies. Recently, a class of renormalization schemes,  parametrized by a single real number $C$,  has been defined  and investigated in the frame of the standard perturbation expansions in powers of the coupling.  In the present paper we investigate the $C$-scheme variation of a Borel-improved QCD perturbation series, which implements information about the large-order divergent character of perturbation theory by means of an optimal conformal mapping of the Borel plane. In the new expansions, the powers of the strong coupling are replaced by a set of expansion functions with properties which resemble those of the expanded correlators, having in particular a singular behavior at the origin of the complex coupling plane. On the other hand, the new expansions have a tamed increase at high orders, as  demonstrated by previous studies in the $\overline{\text{MS}}$ renormalization scheme.  Using as examples the Adler function and the hadronic decay width of the $\tau$ lepton, we investigate the  properties of the Borel-improved expansions in the $C$-scheme, in comparison with the standard expansions in the $C$-scheme and the expansions in $\overline{\text{MS}}$. The variation with the renormalization scale and the prescription for the choice of an optimal value of the parameter $C$ are  discussed.  The good large-order behavior of the Borel-improved expansions is proved also in the $C$-scheme, which is a further argument in favor of using them in applications of perturbative QCD at intermediate energies.
\end{abstract}
\maketitle
\vspace{0.5cm} 
\section{ Introduction}

In the standard QCD perturbation theory, the finite-order
approximations of physical quantities  are renormalization-scale
($\mu$) and renormalization-scheme (RS) dependent. For a truncated expansion of order $N$, the scale and scheme variations, being in principle  ${\cal O}(\alpha_s^{N+1}(\mu))$ corrections, are expected to be  small at  large scales due to asymptotic freedom.   However,  since the perturbative expansion in QCD is a divergent series, with coefficients growing factorially at large orders, the scale and scheme variation might be in practice quite large, especially at intermediate energies where the strong coupling $\alpha_s$ is not very small. 

The quest for in some sense ``optimal'' scale and scheme is important for
meaningful applications. There are several recipes  \cite{PMS, FAC, BLM, Chyla, Brodsky:2011ig, Brodsky:2012rj, Mojaza:2012mf,  Shen:2017pdu}
how to do that. The one proposed in \cite{PMS} and known as the ``principle of
minimal sensitivity'' (PMS) selects the scale and scheme  by the
condition of local scale and scheme invariance. 
Therefore, the PMS  selects the point where the truncated approximant has
locally the property  which the all-order summation must have globally.
A different, process-dependent recipe, known as ``effective charge method'' or ``fastest apparent convergence'' (FAC) criterion  was proposed in \cite{FAC}, while the method advocated in \cite{Brodsky:2011ig, Brodsky:2012rj, Mojaza:2012mf,  Shen:2017pdu}, denoted as  ``principle of maximum conformality'', chooses the scale such as to absorb in the coupling all the nonconformal dependence of the perturbative coefficients. Since the problem is difficult and has so far no generally accepted
solution, perturbative computations are performed mainly
in convenient schemes like the modified minimal subtraction $\overline{{\rm MS}}$ \cite{MSbar}.

Recently, a new class of process-independent renormalization schemes depending on a single real parameter $C$ was defined in \cite{Cscheme}.  In Refs. \cite{Cscheme, Cscheme1, Cscheme2}, the properties of these schemes have been discussed  using the perturbation expansion of the QCD Adler function and the $\tau$ hadronic width, and in \cite{Wu} the class of $C$ schemes was investigated from the point of view of the  maximum conformality  principle \cite{ Brodsky:2011ig, Brodsky:2012rj, Mojaza:2012mf, Shen:2017pdu}. 

 In the present paper, we shall investigate the $C$-scheme in connection with the fact that the perturbation expansions in QCD are divergent series, the expansion coefficients of typical correlators growing factorially at large orders \cite{tHooft, Mueller1985, Mueller1992, Beneke}. This is related to the fact that the QCD correlators  as  functions of  $\alpha_s$ are singular at the origin of the complex $\alpha_s$ plane  \cite{tHooft}, which implies that the  radius of convergence of the expansions in powers of the strong coupling $\alpha_s$ is zero.

Starting from the divergent character of the QCD perturbation series, a modified perturbation expansion was defined in \cite{CaFi1998} and was further investigated in \cite{CaFi2000, CaFi2001, CaFi2009, CaFi_Manchester, CaFi2011, Abbas:2012fi, Abbas:2013} (for a recent review see \cite{Nova}). In this approach, instead of the powers of the strong coupling, a  new set of expansion functions is used,  defined  by means of an optimal (i.e., ensuring the best asymptotic rate of convergence)  conformal mapping of the Borel complex plane.  The properties of the new expansion functions  resemble those of the expanded correlators, by exhibiting in particular a singular behavior at the origin of the  $\alpha_s$ plane. On the other hand, the new expansions have a tamed increase at high orders.   The good convergence properties of the new expansions have been demonstrated  in \cite{CaFi2009, CaFi2011, Abbas:2013} in the $\overline{\text{MS}}$ renormalization scheme on mathematical models that simulate the physical Adler function. 

The aim of the present work is to investigate the properties of these modified perturbation expansions in the class of renormalization schemes defined in \cite{Cscheme}. As in \cite{Cscheme, Cscheme1,Cscheme2},  we use for illustration the Adler function and the hadronic width of the $\tau$ lepton. We start by recalling, in the next section, a few facts about the calculation of these quantities in perturbative QCD.  In Sec. \ref{sec:C} we briefly review, following  \cite{Cscheme}, the $C$-scheme definition of the QCD coupling. In  Sec. \ref{sec:conf} we introduce the modified, nonpower perturbative expansions based on the conformal mapping of the Borel plane, and in Sec. \ref{sec:confC} we rewrite them in  the $C$-scheme.  In Sec. \ref{sec:res} we investigate the $C$-scheme variation of the modified perturbative expansions of the Adler function and the hadronic $\tau$ decay width. We also investigate the large-order behavior of the expansions, using for generating the large-order perturbative coefficients a model of the Adler function proposed in \cite{BeJa}, which we present for completeness in the Appendix.  The last section contains a summary and our conclusions. 
\section{Adler function and $\tau$ hadronic width in perturbative QCD}\label{sec:Adler}
  We recall that the Adler function is the logarithmic derivative of the invariant amplitude of the two-current correlation tensor, $D(s)=-s \,d\Pi(s)/ds$, where $s$ is the momentum squared. As in Ref. \cite{Cscheme} we shall consider the reduced function $\widehat D(s)$ defined as:
\beq\label{eq:D}
\widehat{D}(s) \equiv 4 \pi^2 D(s) -1.
\eeq
From general principles of field theory, it is known that $\wh D(s)$ is an analytic function of real type [i.e., it satisfies the  Schwarz reflection property $\wh D(s^*)=\wh D^*(s)$) in the complex $s$ plane cut along the timelike axis for $s\ge 4 m_\pi^2$.

 At large spacelike momenta $s<0$, the function $\widehat{D}$ is given by the QCD perturbative expansion  \cite{BeJa}
\beq\label{eq:hatD}
\widehat{D}(a_\mu) =\sum\limits_{n\ge 1} a_\mu^n \,
\sum\limits_{k=1}^{n} k\, c_{n,k}\, (\ln (-s/\mu^2))^{k-1},
\eeq
where $a_\mu \equiv \alpha_s(\mu^2)/\pi$ is the renormalized strong coupling in a certain RS at an arbitrary scale $\mu$. As in \cite{Cscheme}, we emphasize  from now on the fact that $\widehat{D}$ is a function  of the QCD coupling $a_\mu$, the dependence on the momentum squared $s$ being implicit.

The leading coefficients  $c_{n,1}$ in (\ref{eq:hatD}) are obtained from the calculation of  Feynman diagrams, while the remaining ones, $c_{n,k}$ with $k>1$ are obtained in terms of  $c_{m,1}$ with $m< n$  and the coefficients $\beta_n$ of the $\beta$ function, which governs the variation of the QCD coupling with the scale  in each RS:
 \begin{equation}\label{eq:rge}
 -\mu\frac {d a_\mu}
{d\mu}\equiv \beta(a_\mu)=\sum_{n\ge 1}
\beta_n a_\mu^{n+1}\,. \end{equation}
We recall that in mass-independent renormalization schemes the first two coefficients  $\beta_1$ and $\beta_2$ are scheme invariant, depending only on the number $n_f$ of active flavors, while  $\beta_n$ for $n\ge 3$ depend on the renormalization scheme. 
In $\overline{{\rm MS}}$, the known coefficients  for $n_f=3$ are (cf. \cite{BCK16} and references therein):
\be\label{eq:betaj}
 \beta_1=\frac{9}{2},\, \beta_2=8,\, \beta_3= 20.12,\,\beta_4=54.46,\,\beta_5=268.16.
\ee

The Adler function  was calculated in $\overline{{\rm MS}}$ to order $\alpha_s^4$, which makes it one of the most precisely known Green functions in QCD. For $n_f=3$ the leading coefficients $c_{n,1}$  have the values (cf. \cite{BCK08} and references therein):
\be\label{eq:cn1}
c_{1,1}=1,\,\, c_{2,1}=1.640,\,\, c_{3,1}=6.371,\,\, c_{4,1}=49.076.
\ee
In the applications done in \cite{Cscheme, Cscheme1, Cscheme2}, an additional term was included, $c_{5,1}=283$, based on the estimate made in \cite{BeJa}, and we shall adopt this value in the present work.

 We shall consider also the perturbative expansion of the total $\tau$ hadronic width. The central
observable is the ratio $R_\tau$ of the total hadronic branching fraction to the electron branching fraction, which can be expressed as
\be\label{eq:Rtau}
R_\tau= 3 \,S_{\rm EW} (|V_{ud} |^2 + |V_{us}|^2 ) (1 + \delta^{(0)} + \ldots),
\ee
where $S_{\rm EW}$ is an electroweak correction, $V_{ud}$ and $V_{us}$ are Cabibbo–Kobayashi–Maskawa (CKM) matrix elements, and $\delta^{(0)}$ is the perturbative QCD contribution. As shown in   \cite{Narison, Braaten, BrNaPi, Diberder}, $\delta^{(0)}$ can be expressed, using analyticity, by an integral involving the values of the Adler function in the complex $s$ plane. In our normalization, this relation is \cite{BeJa}: 
\be\label{eq:delta0}
\hspace{-0.cm}\delta^{(0)} =  \frac{1}{2\pi i} \oint\limits_{|s|=m_\tau^2}\, \frac{d s}{s} \left(1-\frac{s}{m_\tau^2}\right)^3\,\left(1+\frac{s}{m_\tau^2}\right) \wh D(a_\mu). \hspace{-0.15cm}\ee

For the evaluation of  $\delta^{(0)}$  one can either insert in the integral (\ref{eq:delta0}) the expansion   (\ref{eq:hatD}) at a fixed scale and perform the integration of the coefficients with respect to $s$ along the circle, which gives in particular for $\mu=m_\tau$:
\begin{equation}
\label{eq:del0}
\delta_{\rm FO}^{(0)} =
a_\mu + 5.2\,a_\mu^2 + 26.37\,a_\mu^3 + 127.1\,a_\mu^4 + 873.8\, a_\mu^5\ldots
\end{equation}
Alternatively, as proposed in \cite{Diberder}, one can take the variable scale $\mu^2=-s$ in  (\ref{eq:hatD}) and insert in the integral (\ref{eq:delta0})  the renormalization-group improved expansion
\beq\label{eq:hatDCI}
\widehat{D}(a_\mu) =\sum\limits_{n\ge 1} c_{n,1} a_\mu^n \,,\quad \quad \mu^2=-s,
\eeq
with the running coupling calculated by the numerical integration of the renormalization-group equation (\ref{eq:rge}) along the circle $|s|=m_\tau^2$, starting from the spacelike point $s=-m_\tau^2$. These alternatives\footnote{Another approach, proposed in \cite{Ahmady}, includes all the terms available from renormalization-group  invariance and can be expressed as an effective expansion in  powers of the one loop solution of Eq. (\ref{eq:rge}). This summation was investigated in the case of the Adler function in \cite{Abbas:2012py, Abbas:2012fi}.} are knows as  ``fixed-order perturbation theory'' (FOPT) and ``contour-improved perturbation theory'' (CIPT). As remarked first in \cite{BeJa}, contrary to the naive expectations, at the scale set by $m_\tau$ the difference between the predictions of these two summation procedures increased when an additional, five loop term, calculated in \cite{BCK08}, was included in the expansion of the $\tau$ hadronic width. 

The fixed-order series  (\ref{eq:hatD}) is expected to have a poor convergence for $s$ near the timelike axis, where the $s$-dependent expansion coefficients become quite large. 
However, fortuitous cancellations of contributions to the integral  (\ref{eq:delta0}) and the suppressing effect of the weight function in the integrand  might favor the fixed-order series, leading to better results for $\delta^{(0)}$ calculated in FOPT than in CIPT.
Many studies have been devoted to the difference between FO and CI summations, including the analysis of specific models for the Adler function and of the practical implications on the extraction of $\alpha_s(m_\tau^2)$ from data on  hadronic $\tau$ decays (see \cite{BeJa, CaFi2009, CaFi_Manchester, CaFi2011, Nova, Menke, Pich2010, Davier, DeMa, Pich_Manchester, Pich_Muenchen, Pich_2013, Abbas:2012fi, BBJ, Abbas:2013,  Pich-2016, BGMP-2017} and the references therein).

\section{\bf The $C$-scheme QCD coupling}\label{sec:C}

As discussed in \cite{Cscheme}, one can define a new coupling $\hat{a}_\mu$ by using the relation
\bea
\frac{1}{\hat a_\mu} +&& \frac{\beta_2}{\beta_1} \ln\hat a_\mu- \beta_1 \frac{C}{2}
\equiv \beta_1  \ln\frac{\mu}{\Lambda} \nonumber \\&&= \frac{1}{a_\mu} + \frac{\beta_2}{\beta_1} \ln a_\mu -\beta_1 \!\int\limits_0^{a_\mu}\,
\frac{{\rm d}a}{\bar\beta(a)} 
\label{eq:ahat}
\eea
where $\Lambda$ is the scale-invariant QCD parameter and 
  \be
\frac{1}{\bar\beta(a)} \,\equiv\, \frac{1}{\beta(a)} - \frac{1}{\beta_1 a^2}
+ \frac{\beta_2}{\beta_1^2 a} \,.
\ee
From (\ref{eq:ahat}) it is seen that $C$ incorporates the effects of all scheme-dependent terms $\beta_n$  with $n\ge 3$, contained in the function $\bar\beta(a)$. This relation implies also that the scale dependence of $\hat a_\mu$ is given by 
\be\label{eq:betahat}
-\,\mu\,\frac{{\rm d}\ah_\mu}{d\mu} \,\equiv\, \hat\beta(\ah_\mu) \,=\,
\frac{\beta_1 \ah_\mu^2}{\left(1 - \frac{\beta_2}{\beta_1}\, \ah_\mu\right)},
\ee  
and involves only the scheme-independent coefficients $\beta_1$ and $\beta_2$.  Furthermore, as shown in \cite{Cscheme1, Cscheme2}, the $C$ dependence of the coupling $\ah_\mu$ is governed by the same scheme-independent function $\hat\beta$.

Given the coupling $a_\mu$ in a definite RS  at a scale $\mu$,  one can find from (\ref{eq:ahat})  the coupling $\hat a_\mu$ in the $C$-scheme at the same scale  and a definite value of $C$. In order to solve numerically the equation,  is convenient to write it as
\be\label{eq:ahat1}
\frac{1}{\hat a_\mu} + \frac{\beta_2}{\beta_1} \ln\hat a_\mu =f (a_\mu, C),
\ee
where $f(a_\mu, C)$ is a calculable function depending on the coupling $a_\mu$ and the constant $C$. The left hand side of (\ref{eq:ahat1}) is a convex function of  $\hat a_\mu$, exhibiting a single minimum equal to 0.7549 at   $\hat a_\mu=0.5625$, a steep increase towards small values of  $\hat a_\mu$ and a slow logarithmic increase at large  $\hat a_\mu$. Therefore, for values of $a_\mu$ and $C$ such that $f (a_\mu, C)>0.7549$, the equation (\ref{eq:ahat1}) has a unique solution of interest, given by the intersection of the left branch of the function in the l.h.s. with the horizontal line at coordinate $f (a_\mu, C)$.
 
\begin{figure}[htb]
\includegraphics[scale=0.30]{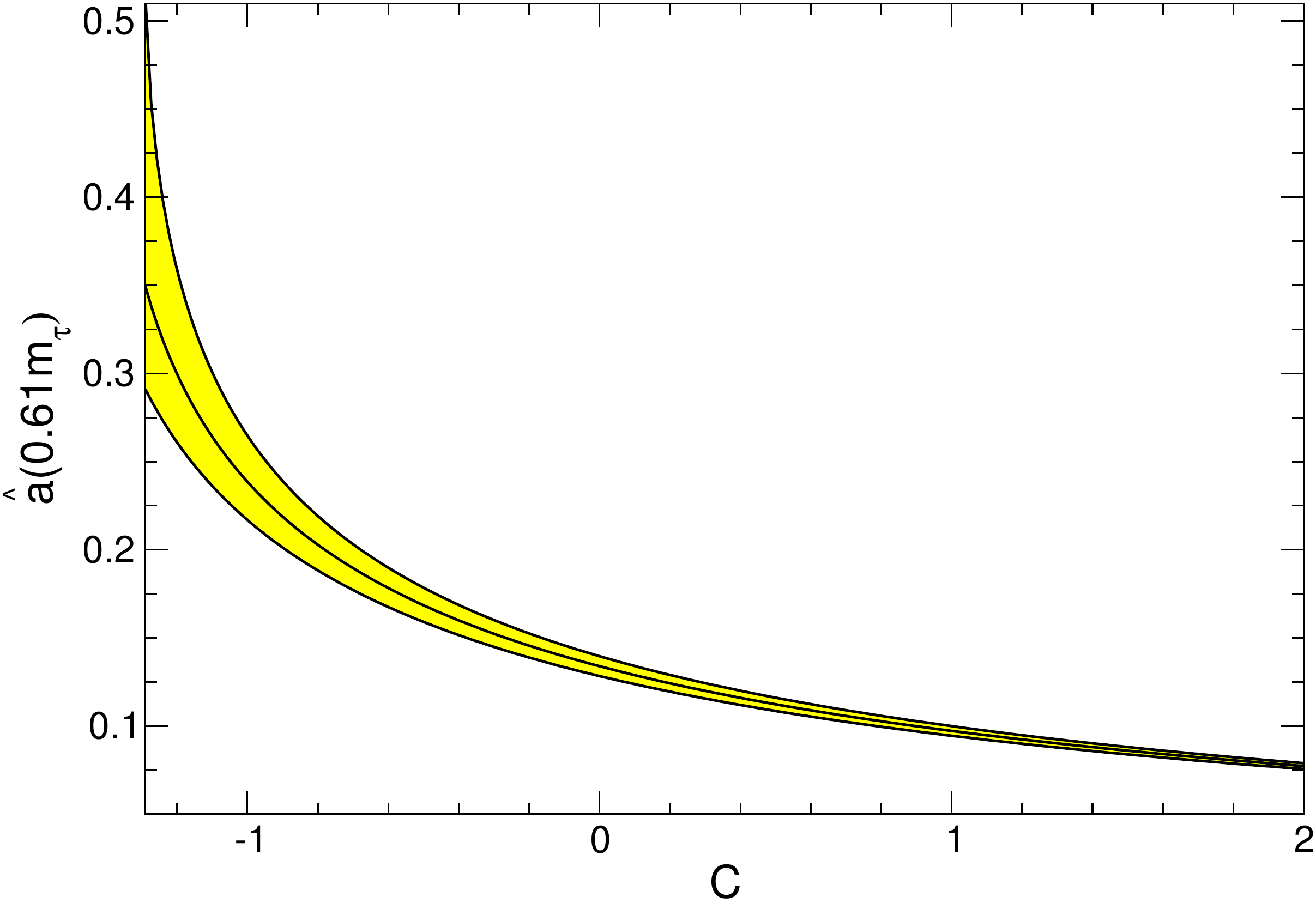}
\caption{The coupling $\hat a_\mu$ found by solving Eq. (\ref{eq:ahat1}) for the scale
 $\mu=0.61 m_\tau$ and $a_\mu= 0.442(20)/\pi$ in the $\overline{{\rm MS}}$ scheme, as a function
of C.   The yellow band corresponds to the $\alpha_s$ uncertainty. \label{fig:0}}
\end{figure}

 As seen from Fig. 1 of \cite{Cscheme}, where the numerical solution is displayed for the input $\mu=m_\tau$ and $a_\mu=0.316(10)/\pi$ in the $\overline{{\rm MS}}$ scheme, the coupling  $\hat a_\mu$ is a decreasing function of $C$. For further illustration we present in Fig. \ref{fig:0} the $C$-dependence of the coupling for the scale $\mu=0.61 m_\tau$, of interest for the analysis performed in Sec. \ref{sec:res}. The coupling $a_\mu= 0.442(20)/\pi$ in the $\overline{{\rm MS}}$ scheme was obtained by solving the RG equation with the input at $\mu=m_\tau$. In this case, solutions of Eq. (\ref{eq:ahat1}) exist only for $C>-1.293$, and one can see from Fig. \ref{fig:0}  the large values of $\hat a_\mu$ near this limit of the validity region.

 From (\ref{eq:ahat}) one can obtain also the perturbative relations between the coupling $a_\mu$ in a certain RS and the coupling $\ah_\mu$  in the $C$-scheme:
\be\label{eq:atoahat} 
\ah_\mu(a_\mu)= \sum_{n\ge 1} \xi_n(C)\, a_\mu^n,\quad a_\mu(\ah_\mu)= \sum_{n\ge 1}\bar\xi_n(C)\, \ah_\mu^n.
\ee 
The explicit forms of these expansions for the $\overline{{\rm MS}}$ scheme are given in  Eqs. (7) and (8) of \cite{Cscheme}. From the comparison with the full solution, found numerically from (\ref{eq:ahat1}),   one can establish the range of $C$ where the perturbative expansions (\ref{eq:atoahat}) are valid. For instance, as shown in \cite{Cscheme}, for $\mu=m_\tau$, when $a_\mu=0.316/\pi$,  the perturbative expansion breaks near $C=-2$. For higher values of the scales, when the coupling $a_\mu$ is smaller, the region of validity of (\ref{eq:atoahat}) is extended to larger negative values of $C$, while for smaller scales the  perturbative expansion breaks at negative values of $C$ closer to the origin.

The perturbative expansions  in the $C$-scheme of the  Adler function $\wh D$ and of the $\tau$ hadronic width have been investigated in  \cite{Cscheme, Cscheme1, Cscheme2}. For instance, the series (\ref{eq:hatDCI}) is rewritten as a series in powers of the $C$-scheme coupling $\ah_\mu$ with $C$-dependent coefficients as
\be\label{eq:hatDC}
\wh D(\ah_\mu) = \sum_{n\ge 1} \hat c_{n, 1}(C)\, \ah_\mu^n,\quad \quad \mu^2=-s.
\ee
Using in (\ref{eq:hatDCI}) the coefficients (\ref{eq:cn1})  and the estimate $c_{5,1}=283$, this expansion reads \cite{Cscheme} 
\bea\label{eq:hatDC1}
&&\wh D(\ah_\mu)= \ah_\mu + (1.64 + 2.25 C)\,\ah_\mu^2 \nn \\
&& \hspace{0.5cm} +\, (7.68 + 11.38 C + 5.06 C^2)\,\ah_\mu^3 \nn \\
&& \hspace{0.5cm} +\, (61.06 + 72.08 C + 47.40 C^2 + 11.4 C^3)\,\ah_\mu^4 \nn\\
&& \hspace{0.5cm}+ \,(348.5 + 677.7 C + 408.6 C^2 + 162.5 C^3 \nn\\
&& \hspace{0.5cm} + \, \,25.6 C^4)\,\ah_\mu^5 + \ldots
\eea
In the same way, from (\ref{eq:del0}) it follows that the expansion of $\delta_{\rm FO}^{(0)}$ in powers of the $C$-scheme coupling $\ah_\mu$ at the scale $\mu=m_\tau$ is \cite{Cscheme}
\bea
\label{eq:del0ah}
&&\hspace{0.cm}\delta_{\rm FO}^{(0)}(\ah_\mu) = \ah_\mu + (5.20 + 2.25 C)\,\ah_\mu^2 \nn \\
 &&\hspace{0.6cm}+\, (27.7 + 27.4 C + 5.1 C^2) \,\ah_\mu^3\nn \\
&&\hspace{0.6cm} +\,(148.4 + 235.5 C + 101.5 C^2 + 11.4 C^3)\,\ah_\mu^4 \nn \\
&&\hspace{0.6cm} +\,(789.6 + 1754.4 C + 1240.4 C^2 + 324.8 C^3  \nn \\
&&\hspace{0.6cm} +\,\, 25.6 C^4) \,\ah_\mu^5+ \ldots 
\eea

The studies performed in \cite{Cscheme, Cscheme1, Cscheme2} showed that the predictions of the perturbation expansions in the $C$-scheme are comparable to those in the $\overline{{\rm MS}}$ scheme. In particular, the difference between the FOPT and CIPT predictions for $\delta^{(0)}$ is not resolved by the $C$-scheme, and the higher-order divergence of the expansions manifests itself to the same or even to a larger extent. 

The works quoted above investigated the scheme dependence of the standard perturbation expansions of the QCD correlators in powers of the strong coupling. In this work we shall investigate the properties of the $C$-scheme using a modified QCD perturbative expansion, defined in \cite{CaFi1998, CaFi2009, CaFi2011}, which we briefly review in the next section.
 
\section{\bf Perturbation expansions with tamed high-order behavior}\label{sec:conf}

The starting point is the remark that the perturbation expansion (\ref{eq:hatDCI}) of the Adler function has a zero radius of convergence,  the  coefficients $c_{n,1}$  increasing like $n !$ at large $n$.
It is convenient to define the Borel transform by 
\be\label{eq:B}
 B(u)= \sum_{n=0}^\infty  b_n\, u^n,\quad\quad  b_n= \frac{c_{n+1,1}}{\beta_0^n \,n!}\,,
\ee
where we used the standard notation $\beta_0=\beta_1/2$. From (\ref{eq:B}) one can derive the formal Laplace-Borel integral representation 
\be\label{eq:Laplace}
\wh D(a_\mu)=\frac{1}{\beta_0} \,\int\limits_0^\infty  
\exp{\left(\frac{-u}{\beta_0 a_\mu}\right)} \,  B(u)\, d u\,.
\ee

As it is known \cite{Mueller1985, Beneke}, the large-order increase of the coefficients of the perturbation series is encoded  in the singularities of the Borel transform  $B(u)$ in the complex $u$ plane. In the particular case of the Adler function,  $B(u)$ has singularities on the semiaxis $u\ge 2$, denoted as  infrared (IR) renormalons, and for $u\le -1$, denoted as ultraviolet (UV) renormalons. The names indicate the regions in the Feynman integrals responsible for the appearance of the corresponding singularities.  Other singularities, at larger values on the positive real axis, are due to specific field configurations known as instantons.  Apart from  the two cuts along the lines $u\geq 2$ and $u\leq -1$, it is assumed that no other singularities are  present in the complex $u$ plane \cite{Mueller1985}. 

 Due to the singularities of $ B(u)$ for $u\ge 2$, the Laplace-Borel integral (\ref{eq:Laplace}) is not defined and requires a regularization.  Several prescriptions have been adopted for various QCD correlators, in particular the  principal value (PV) prescription defined as \cite{Mueller1992, Beneke}
\bea\label{eq:PV}
& \frac{1}{\beta_0} {\rm PV}\,\int\limits_0^\infty  
e^{-\frac{u}{\beta_0 a_\mu}} \,  B(u)\, d u \\
&\equiv \frac{1}{2\beta_0}\left[\int\limits_{C_+}\!\! e^{-\frac{u}{\beta_0 a_\mu}} B(u) d u +\!\! \int\limits_{C_-}\!\! e^{-\frac{u}{\beta_0 a_\mu}} B(u) d u\right],\nonumber
\eea
where $C_\pm$ are lines parallel to the positive real axis, slightly displaced above (below) it. 
 As discussed in \cite{CaNe}, this prescription  is  suitable  from the point of view of the momentum-plane analyticity properties that must be satisfied by QCD correlators like the Adler function.  In particular, it preserves Schwarz reflection principle $\wh D(s^*)=\wh D^*(s)$ in the complex $s$ plane and leads to real values for the Adler function on the spacelike axis $s<0$.

Of course, even after choosing a proper regularization,  an uncertainty  $\delta_{\rm ren}$ related to the renormalon treatment still remains. A natural choice for this uncertainty is the difference between two regularization prescriptions, which can be taken as 
\bea\label{eq:errorPV}
&\frac{1}{\beta_0} \delta_{\rm ren}\{\int\limits_0^\infty  
e^{-\frac{u}{\beta_0 a_\mu}}  B(u)\, d u \}\\
&\equiv \frac{1}{2i\beta_0}\left[\int\limits_{C_+}\!\!e^{-\frac{u}{\beta_0 a_\mu}} B(u) d u -\!\! \int\limits_{C_-}\!\!e^{-\frac{u}{\beta_0 a_\mu}} B(u) d u\right].\nonumber
\eea
This prescription has been used for assessing the uncertainty  of the model for the Adler function proposed  in \cite{BeJa} and reviewed in the Appendix, and we shall use it for some of the expansions investigated in this work.

The singularities of $ B(u)$ set a limitation on the convergence region of the power expansion (\ref{eq:B}): this series converges only inside  the circle $|u|=1$ which passes through the nearest singularity, namely the first UV renormalon. 
As it is known, the domain of convergence of a power series can be increased by expanding in powers of another variable,  which  performs the conformal mapping of the complex plane of the original variable (or a part of it) onto a disk. 

The conformal mapping method was  introduced  in particle physics in  \cite{CiFi, Frazer, CCF} for improving the convergence of the power series used  for the representation of scattering amplitudes.  The new series converges in a larger region,  well beyond the disk of convergence of the original expansion, and also has an increased  asymptotic convergence rate at points lying inside this disk. An important result proved in  \cite{CiFi, CCF}  is that the asymptotic convergence rate is maximal if the new variable maps the entire holomorphy domain of the expanded function onto the unit disk (for a detailed proof see \cite{CaFi2011}). This particular variable is known in the literature as the ``optimal conformal mapping''. 

For QCD,  the method of conformal mapping  is not applicable to the 
formal perturbative series of $\wh D$ in powers of  $a_\mu$, because $\wh D$  is singular 
at the origin of the coupling plane\footnote{In the so-called "order-dependent" conformal mappings, which have been used for expansions in powers of the coupling \cite{ZJ, ZJ1}, the singularity is  shifted away from the origin by a certain amount at each finite perturbative order, and tends to the origin only when an infinite number of terms are considered.}.  However, the method can be applied in a straightforward way to the Borel transform $B(u)$, which is holomorphic in a region containing the origin  $u = 0$  of the Borel complex plane and  can be expanded in powers of  the Borel variable as in (\ref{eq:B}).

The use  of a conformal mapping of the Borel plane was suggested in
\cite{Mueller1992} in order to reduce or eliminate the ambiguities (power 
corrections) due to the large momenta in the Feynman integrals.  This is achieved by a variable that maps on a unit disk the $u$ complex plane cut only along the line $u\leq -1$. 
 As shown however in \cite{CaFi1998}, the conformal mapping proposed in \cite{Mueller1992} 
(and used further in  \cite{Alta, SoSu}) is not optimal in the sense defined above. The optimal mapping, which ensures 
the  convergence of the corresponding power series in the entire doubly-cut Borel plane,  is given by the function \cite{CaFi1998}
\begin{equation}\label{eq:w}
\tilde w(u)=\frac{\sqrt{1+u}-\sqrt{1-u/2}}{ \sqrt{1+u}+\sqrt{1-u/2}}.
\end{equation}

One can check that the function $\tilde w(u)$  maps the complex  $u$ plane cut along the real axis for $u\ge 2$ and $u\le -1$ onto the interior
of the circle $\vert w\vert\, =\, 1$ in the complex plane $w\equiv \tilde w(u)$,  such that  the origin $u=0$ of the $u$ plane
corresponds to the origin $w=0$ of the $w$ plane, and the upper (lower) edges of the cuts are mapped onto the upper
(lower) semicircles in the  $w$ plane. 
By the  mapping (\ref{eq:w}), all  the singularities of the Borel transform, the  UV and IR  renormalons, are pushed on the boundary of the unit disk in the $w$  plane, all at equal distance from the origin. Therefore, the expansion of $B(u)$ in powers of the variable $w\equiv \tilde w(u)$: 
\be\label{eq:Bw}
B(u)=\sum_{n\ge 0} c_n \,w^n, \quad\quad w = \tilde w(u),
\ee
converges in a larger domain that the original series (\ref{eq:B}). 

The expansion can be further improved by exploiting also the fact that the  nature of the leading singularities of $B(u)$ in the Borel plane is known:  near the first branch points, $u=-1$ and $u=2$,  $B(u)$ behaves like
\begin{equation}\label{eq:gammapowers}
 B(u) \sim \frac{r_1}{(1+u)^{\gamma_{1}}} \quad{\rm and} \quad  B(u)  \sim \frac{r_2}{(1-u/2)^{\gamma_{2}}}, 
\end{equation}
respectively, where the residues $r_1$ and $r_2$ are not known, but the exponents
$\gamma_1$ and $\gamma_2$ have been calculated using renormalization-group
invariance \cite{Mueller1992, BBK, BeJa}. For $n_f=3$, their values are  
\begin{equation}\label{eq:gamma12}
\gamma_1 = 1.21,    \quad\quad   \gamma_2 = 2.58 \,. 
\end{equation}

 The knowledge of the nature of the first IR renormalon of the Adler function was exploited for the first time in \cite{SoSu}, where the Borel transform was multiplied by the factor $(1-u/2)^{\gamma_{2}}$ and the product was expanded as a power series\footnote{A similar procedure to account for the first IR renormalon was used in \cite{Pineda, Hoang} for other QCD correlators.}.  But the multiplication with other factors is possible.  
As discussed in \cite{CaFi2009, CaFi2011}, while the optimal conformal mapping (\ref{eq:w}) is unique, there is no unique prescription to implement the knowledge provided by (\ref{eq:gammapowers}). For instance, using the fact that
\be
(1+w)^{2\gamma_1}\sim (1+u)^{\gamma_1}, \quad (1-w)^{2\gamma_2}\sim (1-u/2)^{\gamma_2},
\ee
for $u$ near -1 and 2,  respectively, one can construct these factors in terms of the variable $w$ defined in (\ref{eq:w}).

It is easy to see that, although the product of $B(u)$ with these suitable factors is finite at $u=-1$ and $u=2$, it  still has singularities (branch-points)  at these points, generated by the terms of $B(u)$ which are holomorphic there.
So, the procedure does not eliminate the singularities, but only makes them milder. This is why the procedure was denoted in \cite{SoSu} as ``singularity softening'' and the factors used to multiply the Borel function are referred to as ``softening factors''.   

Since the product of $B(u)$ with the softening factors has still branch-points at $u=-1$ and $u=2$, the optimal conformal mapping for the expansion of the product is the function $w$ defined in (\ref{eq:w}).  Using this remark, we shall adopt the expansion 
\be\label{eq:Bw1}
 B(u)=\frac{1}{(1+w)^{2\gamma_{1}} (1-w)^{2\gamma_{2}}} \sum_{n\ge 0} d_n\, w^n,
\ee
 proposed in \cite{CaFi2009} and shown in  \cite{CaFi2011}  on mathematical models to have good convergence properties at high orders. In order to estimate the effect of each softening factor in (\ref{eq:Bw1}), it is instructive to consider also the alternative expansions 
 \be\label{eq:Bw2}
 B(u)=\frac{1}{ (1-w)^{2\gamma_{2}}} \sum_{n\ge 0} d^{'}_n\, w^n,
\ee
\be\label{eq:Bw3}
 B(u)=\frac{1}{(1+w)^{2\gamma_{1}}} \sum_{n\ge 0} d^{''}_n\, w^n,
\ee
in which only the first IR/UV renormalon has been softened, respectively.

Actually, since the softening factors remove the divergencies at $u=-1$ and $u=2$, leaving only mild singularities at these points, one can expand the product in powers of other conformal variables, which take into account only the position of the more distant singularities of $B(u)$. Such mappings have been considered in \cite{CvLe, CaFi2011}. Moreover, in  \cite{CaFi2011} a detailed study of various softening factors and conformal mappings has been performed.

 As an extreme case, an expansion in powers of the original variable $u$ can be used after ''softening'' the first singularities.   To assess in more detail the importance of various factors, we shall consider the expansions 
\be\label{eq:Bir}
B(u)=\frac{1}{(1-u/2)^{\gamma_2}}\sum_{n\ge 0} b^{'}_n u^n,
\ee
\be\label{eq:Buv}
B(u)=\frac{1}{(1+u)^{\gamma_1}}\sum_{n\ge 0} b^{''}_n u^n,
\ee
and
\be\label{eq:Biruv}
B(u)=\frac{1}{(1-u/2)^{\gamma_2}(1+u)^{\gamma_1}}\sum_{n\ge 0} b^{'''}_n u^n,
\ee
where only the first IR/UV renormalons and both are taken into account, respectively.
One can expect that at low orders the effect of including the known behavior near the first renormalons is important, while at high orders the  singularities which remain in the product, even if they are mild, will deteriorate the convergence  unless a conformal mapping is used.

By inserting the expansions (\ref{eq:Bw}), (\ref{eq:Bw1}),  (\ref{eq:Bw2}),  (\ref{eq:Bw3}), (\ref{eq:Bir}) and (\ref{eq:Buv})  in the Borel-Laplace integral (\ref{eq:Laplace}), we can define new perturbative series for the Adler function. When the expansion of $B(u)$ contains a singularity along the integration range, as in (\ref{eq:Bw}), (\ref{eq:Bw1}), (\ref{eq:Bw2}) and (\ref{eq:Bir}), we shall adopt the PV prescription defined in (\ref{eq:PV}).  Thus, for instance, the expansion 
(\ref{eq:Bw1}) leads to
\be\label{eq:DnewCI}
\wh D(a_\mu)=\sum\limits_{n\ge 0} d_n {\cal W}_n(a_\mu), 
\ee
where the expansion functions are
\be\label{eq:Wn}
\hspace{-0.05cm}{\cal W}_n(a_\mu)=\frac{1}{\beta_0}{\rm PV} \int\limits_0^\infty\!  \frac{{\rm e}^{-\frac{u}{\beta_0 a_\mu}}(\tilde w(u))^n}{(1+\tilde w(u))^{2\gamma_{1}} (1-\tilde w(u))^{2\gamma_{2}}}du.
\ee
 We emphasize that the definition  (\ref{eq:DnewCI}) implies the permutation of the summation and the integration, which is not a trivial step, therefore  (\ref{eq:DnewCI}) represents a genuinely new perturbation expansion in QCD. 

The properties of the new expansions like (\ref{eq:DnewCI}) have been investigated in detail in \cite{CaFi2000, CaFi2001, CaFi2011}, and we briefly summarize them here.
By construction, when reexpanded in powers of $a_\mu$, the series (\ref{eq:DnewCI}) reproduces the expansion (\ref{eq:hatDCI}) with the coefficients $c_{n,1}$   known from Feynman diagrams.  The expansion functions ${\cal W}_n(a_\mu)$ are singular at $a_\mu=0$ and have divergent expansions when expanded in powers of $a_\mu$, resembling the expanded function $\wh D(a_\mu)$ itself.  On the other hand, the expansion (\ref{eq:DnewCI}) has a tamed behavior at high orders, and, under certain conditions, it  may even converge in a domain of the $s$-plane. 

Since the expansion functions ${\cal W}_n(a_\mu)$ defined in (\ref{eq:Wn}) are no longer powers of the coupling $a_\mu$,  the new expansion (\ref{eq:DnewCI}) can be viewed as a ``nonpower perturbation theory'' (NPPT) \cite{CaFi2011}. We shall  also refer to it as  ``Borel-improved'' expansion,  to emphasize the fact that (\ref{eq:DnewCI})  is defined by the analytic continuation of the Borel series (\ref{eq:B}) outside the original convergence disk $|u|<1$, to the whole  Borel plane up to its cuts.

We presented above the steps leading to the new expansion (\ref{eq:DnewCI}) starting from the renormalization-group improved expansion (\ref{eq:hatDCI}), but similar steps can be followed starting from the fixed-order expansion (\ref{eq:hatD}), when the scale $\mu^2$ is different from the energy squared $-s$. The explicit formulas are given in \cite{CaFi2011}.  By inserting in  (\ref{eq:Laplace})  the  expansions (\ref{eq:Bw}), (\ref{eq:Bw2}), (\ref{eq:Bw3}), (\ref{eq:Bir}) and (\ref{eq:Buv}) of the Borel transform, we obtain also the alternative expansions of $\wh D$ which will be investigated in  our study.

\section{\bf Borel-improved expansions in the $C$-scheme}\label{sec:confC}
 The nonpower expansions defined  above have been investigated up to now in the $\overline{{\rm MS}}$ renormalization scheme. However,   the construction presented in the previous section is general and can be performed in any renormalization scheme. Starting from the expansion  (\ref{eq:hatDC}) of the Adler function in powers of the $C$-scheme coupling $\ah_\mu$, we define the corresponding Borel transform as
\be\label{eq:BhatC}
\hat B(u, C)= \sum_{n=0}^\infty \hat b_n(C) u^n,\quad\quad \hat b_n= \frac{\hat c_{n+1,1}(C)}{\beta_0^n \,n!}\,,
\ee
 and obtain the formal Laplace-Borel integral representation 
\be\label{eq:LaplaceC}
\wh D(\ah_\mu)=\frac{1}{\beta_0} \,\int\limits_0^\infty  
\exp{\left(\frac{-u}{\beta_0 \ah_\mu}\right)} \, \hat B(u, C)\, d u\,.
\ee

A useful remark  is that the position and the nature of the first singularities of the Borel transform in the $u$ plane depend only on the first two coefficients, $\beta_1$ and $\beta_2$, of the $\beta$ function, which are  scheme-independent \cite{Mueller1992,BBK, BeJa}. It follows that, for every $C$, the first singularities of the function $\hat B(u, C)$  are situated at $u=-1$ and $u=2$, and the nature of the singularities is given by (\ref{eq:gammapowers}). Therefore,  we can use for $\hat B(u, C)$ the expansions similar to those written in (\ref{eq:Bw}) and 
(\ref{eq:Bw1})-(\ref{eq:Biruv}) in the $\overline{{\rm MS}}$ scheme. In particular, we write the expansion
\be\label{eq:BwC}
 \hat B(u, C)=\frac{1}{(1+w)^{2\gamma_{1}} (1-w)^{2\gamma_{2}}} \sum_{n\ge 0} \hat d_n(C)\, w^n,
\ee
similar to (\ref{eq:Bw1}), the only difference being that now the coefficients $\hat d_n$  depend on $C$. By inserting this expansion into (\ref{eq:LaplaceC}), we define the Borel-improved expansion of the Adler function in the $C$-scheme by
\be\label{eq:DnewCIC}
\wh D(\ah_\mu)=\sum\limits_{n\ge 0} \hat d_n(C) \, \hat{\cal W}_n(\ah_\mu), 
\ee
where the expansions functions $\hat{\cal W}_n(\ah_\mu)$ are obtained from (\ref{eq:Wn}) by formally replacing $a_\mu$ with the  $C$-dependent coupling $\ah_\mu$.

 For illustration, we list the first coefficients $\hat d_n(C) $ appearing in  (\ref{eq:DnewCIC}):
\bea\label{eq:dnC}
&&\hspace{-0.4cm}\hat d_0=1,  \,\quad \hat d_1=-0.80 + 2.67\, C,\,\nn  \\&&\hspace{-0.4cm} \hat d_2= 1.33 + 2.46 \,C + 
    3.56\, C^2,\nn  \\&&\hspace{-0.4cm} \hat d_3 = 10.69 + 2.31\, C + 8.15\, C^2 + 
    3.16 \,C^3, \\&& \hspace{-0.4cm}\hat d_4= 1.15 + 23.44\, C + 8.37\, C^2 + 
    11.02\, C^3 + 2.11\, C^4.\nn
\eea
In a similar way, starting from the expansion (\ref{eq:del0ah}) of $\delta_{\rm FO}^{(0)}$ in the $C$-scheme, we can define the improved series in terms of the same set of functions $\hat{\cal W}_n(\ah_\mu)$, since from the definition (\ref{eq:del0}) it follows that the position and the nature of the first singularities of the Borel transform of 
$\delta_{\rm FO}^{(0)}$ are the same as those of the Adler function, So, we can write
\be\label{eq:del0newC}
\delta_{\rm FO}^{(0)}(\ah_\mu)=\sum\limits_{n\ge 0} \hat \delta_n(C) \, \hat{\cal W}_n(\ah_\mu), 
\ee
where $\mu=m_\tau$ and the first coefficients  $\hat \delta_n(C)$ are
\bea\label{eq:delnC}
&& \hspace{-0.4cm}\hat\delta_0=1, \, \quad \hat\delta_1=3.42 + 2.67\, C, \,\nn  \\&&\hspace{-0.4cm} \hat\delta_2=6.62 + 13.72\, C + 
    3.56\, C^2,\nn  \\&&\hspace{-0.4cm}  \hat\delta_3=4.96 + 31.82\, C + 23.16 \,C^2 + 
    3.16\, C^3,  \\&& \hspace{-0.4cm} \hat\delta_4=-14.22 + 29.32\, C + 65.63\, C^2 + 
    24.36\, C^3 + 2.11\, C^4.\nn
\eea

In CIPT, the calculation  involves the numerical integration (\ref{eq:delta0})  of the expansion (\ref{eq:DnewCIC}), written as:
\bea\label{eq:delCIC}
&&\delta_{\rm CI}^{(0)}(\ah_{m_\tau})=\sum\limits_{n\ge 0} \hat d_n(C) \\
&&\times \frac{1}{2\pi i}\,\oint\limits_{|s|=m_\tau^2}\, \frac{d s}{s} \left(1-\frac{s}{m_\tau^2}\right)^3\,\left(1+\frac{s}{m_\tau^2}\right) \hat{\cal W}_n(\ah_\mu),\nn 
\eea
where the coefficients $\hat d_n$ are given in (\ref{eq:dnC}) and the expansion functions $\hat{\cal W}_n$, defined below  (\ref{eq:DnewCIC}), depend on the running coupling $\ah_\mu$ at the scale $\mu^2=m_\tau^2 \exp(i (\phi-\pi))$, $\phi\in (0, 2\pi)$, which is calculated along the circle by integrating the renormalization-group equation  (\ref{eq:betahat}) in the $C$-scheme, starting from a given  value at the scale $\mu=m_\tau$. Therefore, the whole expansion depends only on the $C$-scheme coupling $\ah_{\mu}$ at the scale $\mu=m_\tau$.

\section{Results}\label{sec:res}
\subsection{Adler function}\label{sec:D}

As a  first application, we calculate  the Adler function (\ref{eq:D}) at the spacelike point  $s=-m_\tau^2$ using the Borel-improved expansion in the $C$-scheme.
 Following Ref. \cite{Cscheme}, we take first the scale $\mu=m_\tau$, when the perturbative expansion (\ref{eq:hatD}) writes as the renormalization-group improved series (\ref{eq:hatDCI}). Other scales will be also considered below.  We use the value $\alpha_s(m_\tau^2)=0.316\pm 0.010$ in the  $\overline{{\rm MS}}$ scheme, which follows from the PDG value of the strong coupling at the scale $\mu=m_Z$  \cite{PDG}. The corresponding $C$-dependent value of the coupling $\ah_{m_\tau}$ was obtained by numerically solving  equation (\ref{eq:ahat1}), as explained in Sec. \ref{sec:C}. 

\begin{figure}[htb]
\includegraphics[scale=0.30]{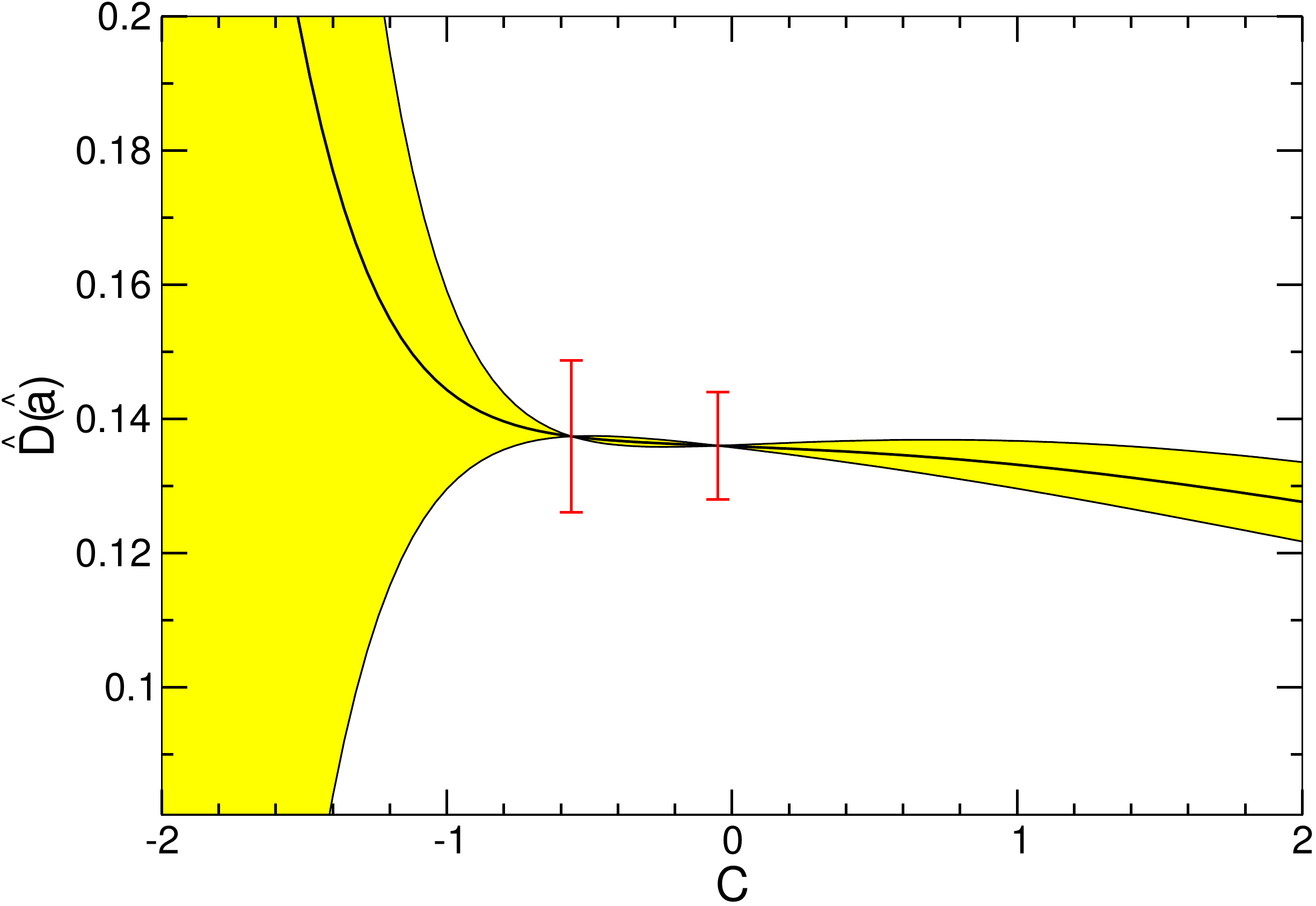}
\caption{Borel-improved expansions in the $C$-scheme of the Adler function $\wh D$ at $s=-m_\tau^2$. The central line is the  expansion to order $\alpha_s^5$. The yellow band is obtained by either removing or doubling the last term. Marked in red are the points where the last term of the expansion vanishes and the magnitude of the last nonvanishing term is shown. \label{fig:1}}
\end{figure}

In Fig. \ref{fig:1}, we show the variation with $C$ of the Borel-improved expansions 
in the $C$-scheme, given in Eqs. (\ref{eq:DnewCIC})-(\ref{eq:dnC}), for $C$ in the range $(-2, 2)$: the central black line represents the  expansion to order $\alpha_s^5$, and the lines delimiting the yellow region are obtained by either removing or doubling the coefficient $\hat d_4$ given in (\ref{eq:dnC}).

The three curves exhibit a plateau where the expansions are stable with respect to the variation of $C$. There are two points, represented in red, where the three expansions coincide: the rightmost point, $C_0=-0.05$,  is the only real solution of the equation $\hat d_4=0$ in the range $(-2,2)$. At the other point, $C_0'=-0.56$,  the expansion function $\hat{\cal W}_4$ itself vanishes, a phenomenon which can take place since the expansion functions are no longer simple powers of the coupling $\ah_\mu$. At these points we indicate the magnitude of the previous nonvanishing term, $\hat d_3\hat{\cal W}_3$.

\begin{figure}[htb]
\includegraphics[scale=0.30]{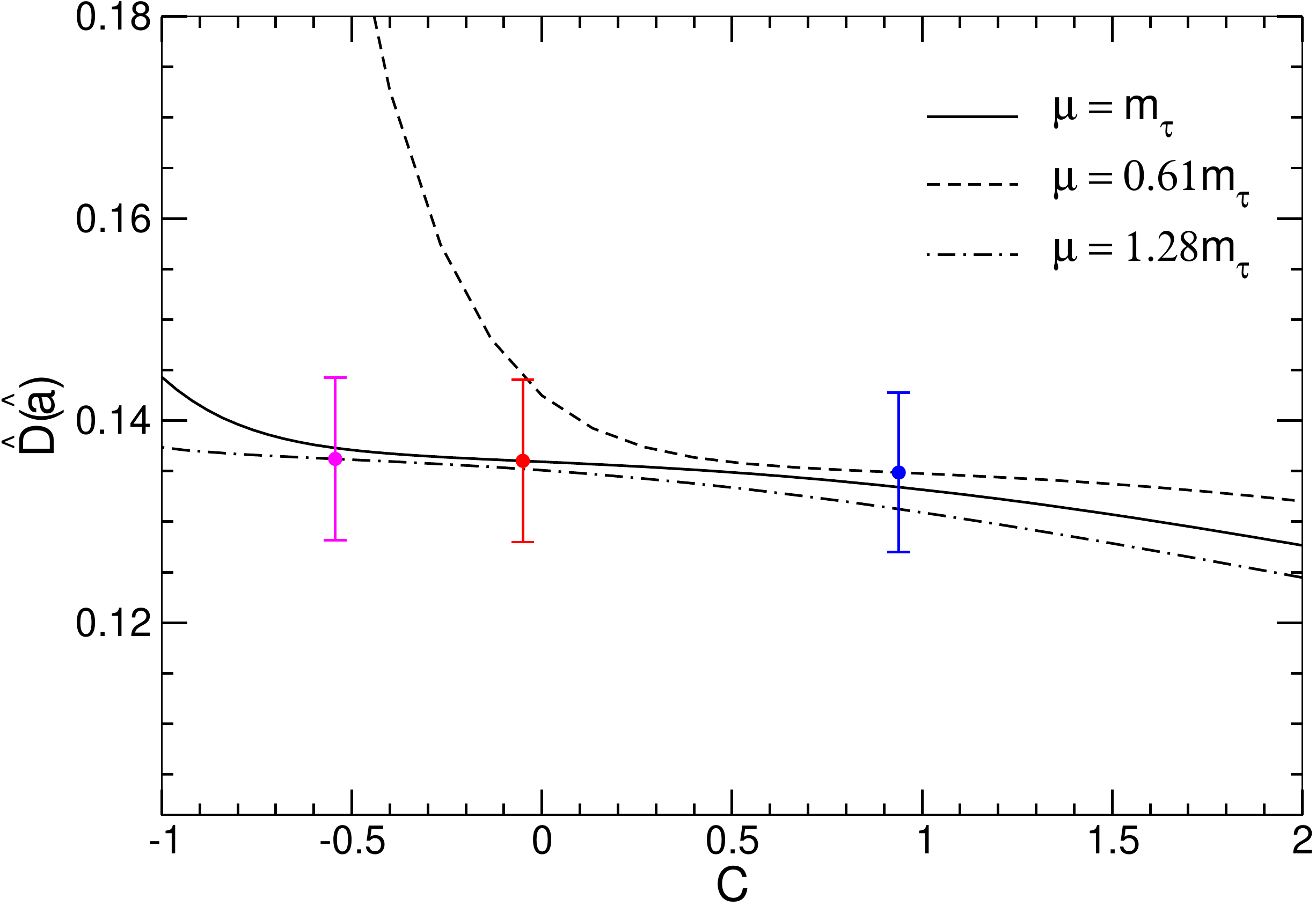}
\caption{ Borel-improved expansions to order $\alpha_s^5$ in the $C$-scheme of the Adler function $\wh D$ at $s=-m_\tau^2$, for the scales $\mu=m_\tau$,  $\mu=0.61 m_\tau$ and $\mu=1.28 m_\tau$, as functions of $C$. We show in each case the optimal points where the last term of the expansion vanishes and the magnitude of the last nonvanishing term. \label{fig:1scale}}
\end{figure}
  In Ref. \cite{Cscheme}, the point where the three expansions coincide has been selected as the optimal value of the parameter $C$. If there are more values of $C$ which satisfy this condition, the optimal one was taken such as to lead to a minimal value for  the last nonvanishing term, which was interpreted as a truncation error\footnote{A bit surprising is the fact (discussed in the last section), that the optimal $C$ and the error prescription  proposed in Ref. \cite{Cscheme} only work at 5 loops.   In fact, the
5th order has not been computed analytically, and a numerical estimate is used.}.  In the present work, we shall consider for the beginning the same prescription, which is in the spirit of the ``fastest apparent convergence'' \cite{FAC} and ''minimum sensitivity'' \cite{PMS} principles.  In our case this point is $C_0=-0.05$, which leads to the central value $\wh D=0.1364$. For comparison, we note that the   Borel-improved expansion to $O(\alpha_s^5)$  in the $\overline{{\rm MS}}$ scheme gives the close value 0.1360, while the standard expansions in the $C$-scheme and in  $\overline{{\rm MS}}$ lead to the values 0.1343 and 0.1316, respectively \cite{Cscheme}.

The main sources of theoretical errors to be attached to these values are the uncertainty of the strong coupling  $\alpha_s(m_\tau^2)$, the effect of truncating the perturbative series at a finite order and the variation of the renormalization scale. The first error can be easily calculated in each case.  The truncation error  was taken  in  \cite{Cscheme} as the magnitude of the last nonvanishing term kept in the expansion.  For the Borel-improved expansions the last nonvanishing terms turn out to be equal to 0.0080 and 0.0030  in the $C$ and $\overline{{\rm MS}}$ schemes, respectively, while for the standard expansions the corresponding values are 0.0070 and 0.0029.  Actually,  no definite way for assessing the effect of the unknown higher-order terms in the perturbative expansion exists. The choice made in  \cite{Cscheme},  in the spirit of asymptotic  expansions, may be affected by numerical accidental cancellations  leading to values too  small for the last coefficient. 

Following  \cite{Cscheme}, we adopted in the above calculation the scale $\mu=m_\tau$, when all the logarithms in the expansion (\ref{eq:hatD}) are summed leading to the renormalization-group improved  series (\ref{eq:hatDCI}). However, it is instructive to investigate also other scales. Following \cite{Davier}, we parametrized the scale dependence by writing $\mu^2=-\xi s$ with $\xi =1\pm 0.63$. A similar range of $\mu$ has been adopted in \cite{Pich_2013}. In particular, for the point $s=-m_\tau^2$ used in the calculation of the Adler function,  the scale is written as $\mu=\sqrt{\xi} m_\tau$, and varies between $0.61 m_\tau$ and $1.28 m_\tau$. 

By comparing the  perturbative expansion (\ref{eq:atoahat}) with the full solution determined numerically from (\ref{eq:ahat1}), one can see that for scales $\mu$ larger than $m_\tau$, when  the coupling $\alpha_s$ is smaller, the perturbative expansion is valid on a larger interval, extending to the left of $C=-2$. On the other hand, as seen from Fig. \ref{fig:0}, for smaller $\mu$ the coupling is larger and the validity of the perturbative expansion (\ref{eq:atoahat}) breaks down at values of $C$ closer to 0. 

In Fig. \ref{fig:1scale} we show the Borel-improved expansions in the $C$-scheme of the Adler function $\wh D$ at $s=-m_\tau^2$,  for the scale $\mu=m_\tau$ and for the extreme values   $\mu=0.61 m_\tau$ and $\mu=1.28 m_\tau$ of the range considered. We show also in each case the points where the last term of the expansion vanishes and the magnitude of the last nonvanishing term.  For $\mu=m_\tau$ the curve and the optimal point coincide with those given in  Fig. \ref{fig:1}.  

In order to quantify the variation with the scale, we can  compare the results for different scales at a fixed value of $C$, namely the optimal $C$ determined for $\mu=m_\tau$  (the point marked in red). A different prescription would be to compare the optimal values for each scale, obtained with the corresponding optimal values of $C$. As seen from Fig. \ref{fig:1scale},  for the Adler function the first definition appears to give reasonable estimates of the errors, while with the second prescription the errors appear to be underestimated. Therefore, for the Borel-improved expansion in the $C$-scheme we adopt the first prescription for the error.  

On the other hand, for the standard expansions in the $C$-scheme, where the optimal value of $C$ for $\mu=m_\tau$ is $C_0=-0.78$ \cite{Cscheme}, the perturbative expansion for $\mu$ towards the lower end of the range is not well behaved. In this case, one can either use for low values of $\mu$ the second prescription mentioned above, or reduce the interval of scale variation.  This ambiguity may induce some distortions in the error estimate, which must be taken with caution in these cases\footnote{The problems encountered when varying the scale indicate that a more elaborate prescription for the optimal $C$ than that proposed in \cite{Cscheme} may  be necessary. We will make some comments on this in the final section.}.  

By including all sources of error, we obtain the result of the Borel-improved expansion in $C$-scheme as
\be\label{eq:D5}
\wh D=0.1364 \pm 0.0080\, ^{+0.0078}_{-0.0012}   \pm 0.0061,
\ee
where the first error is the magnitude of the last nonvanishing term in the expansion, the second is the uncertainty due to scale variation and the third is due to the uncertainty in $\alpha_s(m_\tau^2)$.

For comparison we note that to the same order the Borel-improved expansion  in the $\overline{{\rm MS}}$ scheme gives 
\be
\wh D=0.1360 \pm 0.0030 \,^{+0.0034}_{-0.0010} \pm 0.0061,
\ee  where the significance of the uncertainties is the same. The standard expansion in the $C$-scheme gives   
\be
\wh D= 0.1343\pm 0.0070\,^{+0.0001}_{-0.0016}   \pm 0.0067,
\ee
  where the error due to the scale variation has been estimated from the differences of the optimal values at different scales, and might be underestimated.  For the standard expansion in $\overline{{\rm MS}}$ the result is
\be\wh D=0.1316\pm 0.0029\,^{+0.0029}_{-0.0030} \pm 0.0060.\ee

We note finally that, due to the conformal mapping (\ref{eq:w}) and the softening factors (when present), the definition of the Borel-improved expansions requires a regularization prescription of the Borel-Laplace integral.  For calculating the central values we adopted the PV prescription (\ref{eq:PV}), which has the advantage of preserving Schwarz reflection principle and leads to real values for the Adler function on the spacelike axis. Since the PV prescription is part of the definition of the Borel-improved expansions, we do not include the  regularization ambiguity as an additional error. For completeness, we mention that the definition (\ref{eq:errorPV}) leads to similar values, equal to 0.0098 and 0.0089,  respectively, for the prescription ambiguity of the Borel-improved expansions in the $C$ and   $\overline{{\rm MS}}$ schemes.

\begin{figure}[htb]
\includegraphics[scale=0.30]{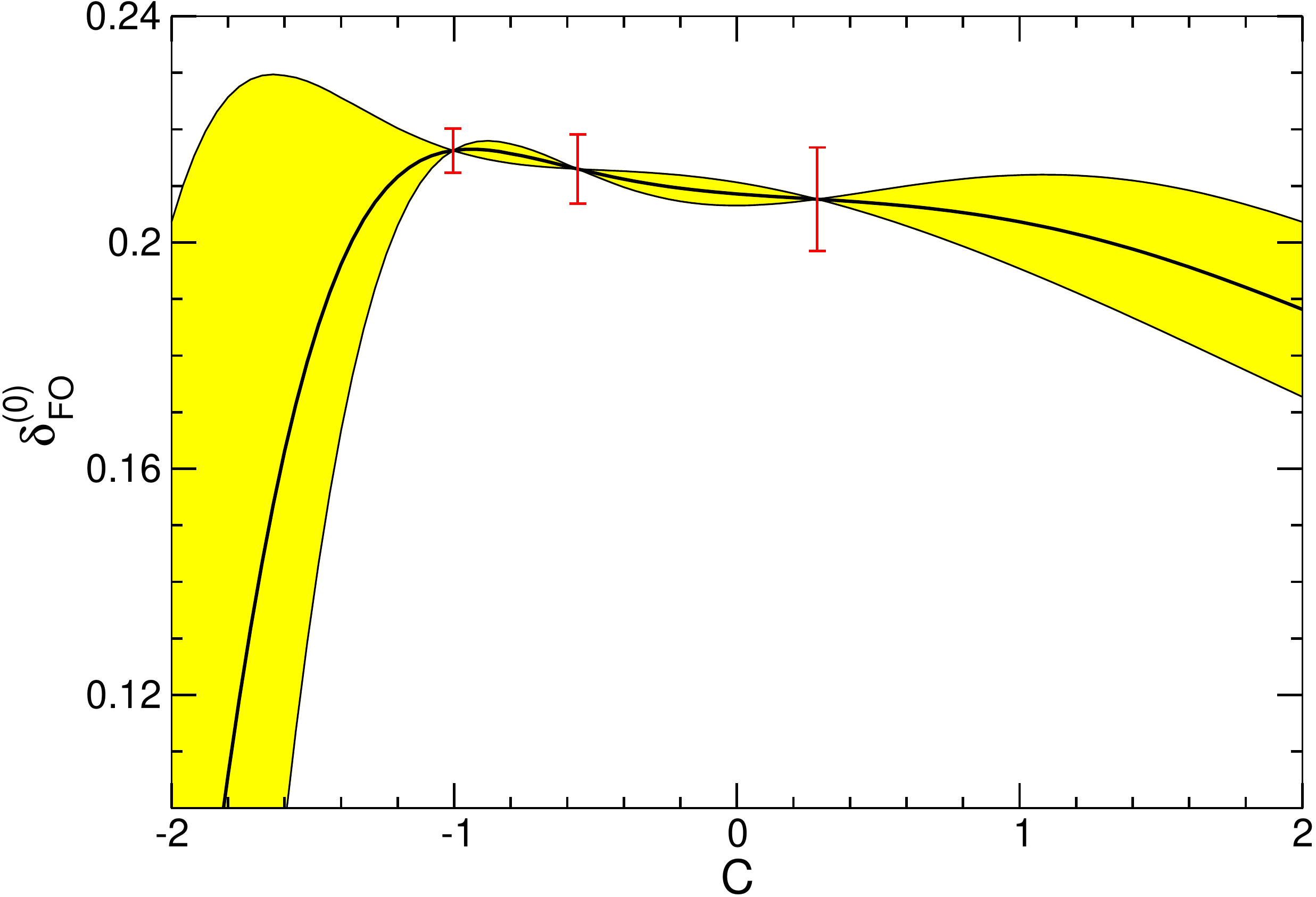}
\caption{Borel-improved expansion of $\delta_{\rm FO}^{(0)}$ to order $O(\alpha_s^5)$ in the $C$ scheme as a function of
$C$. The yellow band arises from either removing or doubling the last
term in the expansion.  In red are marked the points where the last term in the expansion vanishes, and the magnitude of the previous one is indicated. \label{fig:2}}
\end{figure}
\subsection{$\tau$ hadronic width}\label{sec:delta0}

We consider now the calculation of the QCD correction $\delta^{(0)}$ to the $\tau$ hadronic width (\ref{eq:Rtau}). As discussed in Sec. \ref{sec:Adler}, there are two standard summation methods, CIPT and FOPT, which differ essentially by the choice of the renormalization scale in performing the integral (\ref{eq:delta0}) along the circle in the $s$ plane.  The difference between FOPT and CIPT  in the  $\overline{{\rm MS}}$ scheme is the main source of uncertainty in the extraction of the strong coupling from hadronic $\tau$ decays. As discussed in Ref. \cite{Cscheme}, this difference persists also in the $C$ scheme.  
In the present subsection we investigate the problem using the Borel-improved FO and CI expansions of   $\delta^{(0)}$ in the $C$ scheme, discussed in Sec. \ref{sec:confC}.

 The  Borel-improved FO expansion of $\delta^{(0)}$ in the $C$-scheme is given in Eqs. (\ref{eq:del0newC}) and (\ref{eq:delnC}). According to standard practice in the perturbative calculations of $\tau$ hadronic width, the scale was first fixed at $\mu=m_\tau$.  As in the previous subsection, one explored also other scales, taking $\mu^2=-\xi s$ with $\xi=1\pm 0.63$ \cite{Davier}.

 In Fig. \ref{fig:2} we show the Borel-improved FO expansion of $\delta^{(0)}$ in the $C$ scheme to order $\alpha_s^5$ as central line, and the curves  obtained by either doubling or removing the last term in the expansion, which delimit the yellow band.  As seen from the figure, there are three points where the curves coincide: the leftmost and the rightmost ones are the solutions of the equation $\hat\delta_4=0$, while the middle point is the solution of the equation $\hat{\cal W}_4=0$, encountered already in Fig. \ref{fig:1}. It turns out that the magnitude of the last nonvanishing term, $\hat \delta_3 \hat{\cal W}_3$, is minimal for the leftmost point, $C_0=-1$, which we adopt as optimal.  This leads to the prediction
\be\label{eq:del0FOBC}
\delta_{\rm FO}^{(0)}=0.2207 \pm 0.0039 \,^{+0.0003}_{-0.0082} \pm 0.0195,
\ee
where the first error is the magnitude of the last nonvanishing term in the expansion, the second one is obtained from the scale variation and the third accounts for the uncertainty in the coupling. We note that we encountered the situation mentioned above for the Adler function in the standard $C$-scheme, i.e., the optimal $C_0$ for $\mu=m_\tau$ is close to -1, where the expansions at low scales $\mu$ are not well-behaved. Therefore, the errors due to scale variation quoted in (\ref{eq:del0FOBC}), calculated by taking the differences of the optimal values for each scale, might be underestimated.

For comparison, the Borel-improved expansion in the $\overline{{\rm MS}}$ scheme to $O(\alpha_s^5)$ gives 
\be\label{eq:del0FOB}
\delta_{\rm FO}^{(0)}=0.2104 \pm 0.0031\,^{+0.0108}_{-0.0020} \pm 0.0136,\ee
 where the significance of the terms is the same, while
 the standard expansion in the $C$-scheme  predicts 
\be\label{eq:del0FOCst}
\delta_{\rm FO}^{(0)}=0.2047\pm 0.0034\,^{+0.0002}_{-0.0059}\pm  0.0133,\ee and  the standard expansion in the $\overline{{\rm MS}}$ scheme gives  \be\label{eq:del0FOst}\delta_{\rm FO}^{(0)}= 0.1991\pm 0.0061\,^{+0.0042}_{-0.0073}\pm 0.0119. \ee 
We note that for the standard expansion in the $C$-scheme, where the optimal $C$ for the central scale is close to -1  \cite{Cscheme}, the errors due to scale variations have been calculated with the second prescription discussed above, and may be underestimated.
  We mention finally that for the Borel-improved expansions the uncertainty (\ref{eq:errorPV}) due to the regularization prescription  turns out to be 0.0157 for the $C$ scheme and 0.0080 for $\overline{{\rm MS}}$. 

\begin{figure}[htb]
\includegraphics[scale=0.3]{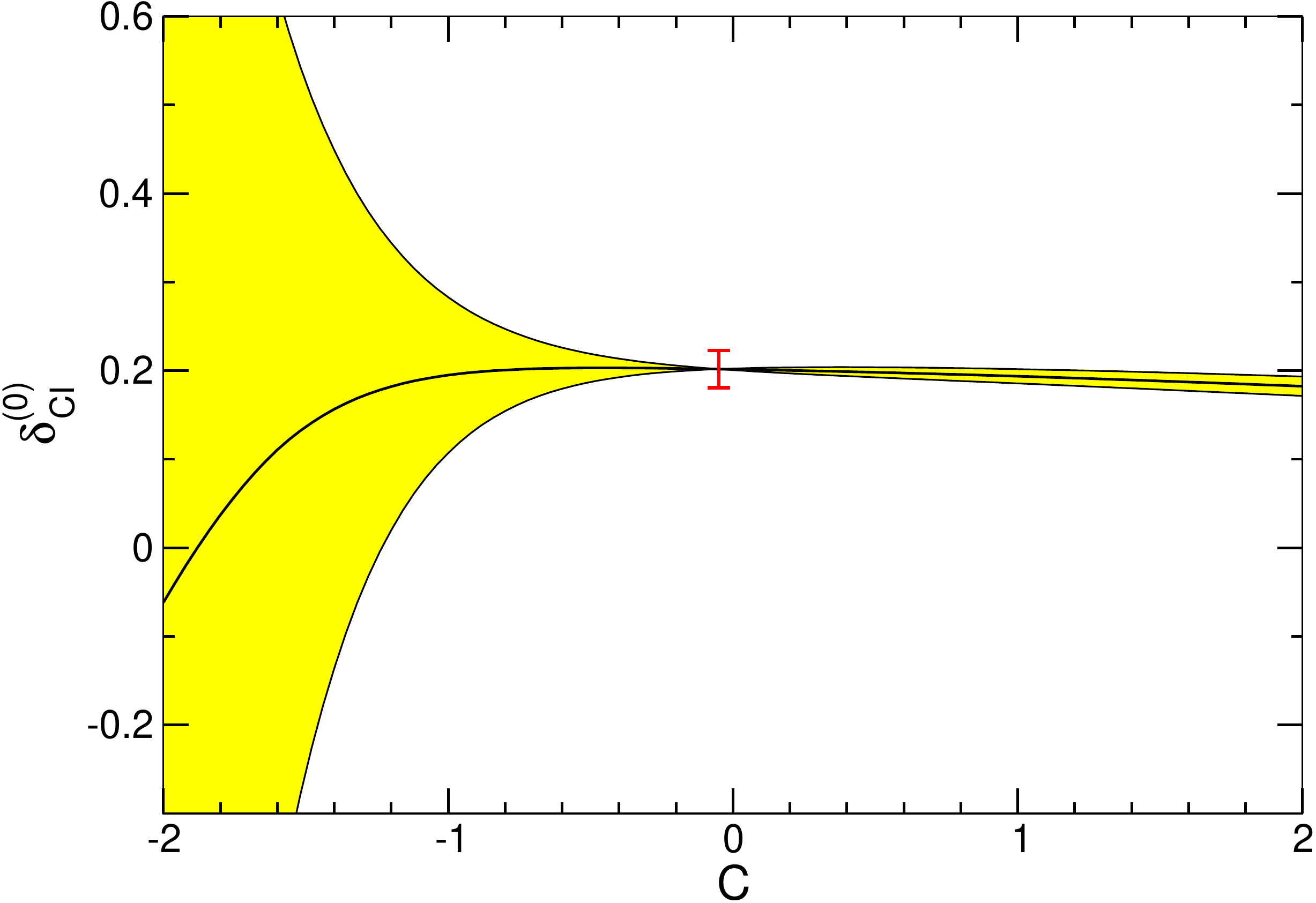}
\caption{Borel-improved expansion of $\delta_{\rm CI}^{(0)}$ to $O(\alpha_s^5)$ in the $C$ scheme as a function of $C$. The yellow band arises from either removing or doubling the fifth-order term. In red is marked the point where the last term in the expansion  vanishes, and the magnitude of the previous term is indicated.  \label{fig:3}}
\end{figure}

For the calculation in CIPT, we use Eq. (\ref{eq:delCIC}) and the numerical procedure described at the end of Sec. \ref{sec:confC}. In Fig. \ref{fig:3} we show the variation with $C$ of the Borel-improved expansion of $\delta_{\rm CI}^{(0)}$ calculated to order $\alpha_s^5$ in the $C$ scheme, and also the curves obtained by either removing or doubling the last coefficient $\hat d_4(C)$. For the scale $\mu^2=-s$ these curves intersect each other at $C_0=-0.05$, where $\hat d_4(C)$ vanishes. By varying also the scale as $\mu^2=-\xi s$, we obtain:
\be\label{eq:del0CIBC}
\delta_{\rm CI}^{(0)}=0.2018\pm 0.0211\,^{+0.0008}_{-0.0139}   \pm 0.0123,
\ee
where, as above, the first error is the magnitude of the last nonvanishing term kept in the expansion, the second is due to scale variation and the third accounts for the uncertainty in the coupling. 
For comparison, the Borel-improved expansion in the $\overline{{\rm MS}}$ scheme gives 
\be\label{eq:del0CIBst}
\delta_{\rm CI}^{(0)}=0.1997 \pm 0.0018\,^{+0.0007}_{-0.0054}\pm  0.0119,\ee while the standard expansion in the $C$-scheme predicts  
\be \label{eq:del0CICst}
\delta_{\rm CI}^{(0)}=0.1840\pm 0.0062 \, ^{+0.0002}_{-0.0044} \pm 0.0084, \ee  and the standard expansion in the $\overline{{\rm MS}}$ scheme gives:
\be\label{eq:del0CIst} \delta_{\rm CI}^{(0)}=0.1826 \pm 0.0032\,^{+0.0004}_{-0.0029} \pm 0.0082.\ee  
We note that for the standard expansion in the $C$-scheme, where the optimal $C$ for the central scale is large and negative \cite{Cscheme}, the errors due to scale variations have been calculated with the second prescription discussed above, and may be underestimated.
The prescription ambiguity for the Borel-improved expansions turns out to be very small in this case, below 0.0011 for both $C$ and  $\overline{{\rm MS}}$ schemes.

From (\ref{eq:del0FOBC}) and (\ref{eq:del0CIBC}) it is seen that the difference between the FOPT and CIPT predictions for $\delta^{(0)}$  persists also for the Borel-improved expansions in the $C$-scheme. On the other hand, one can notice the close results obtained with the Borel-improved CI expansions  and the standard FO expansions,  for both renormalization schemes: for the $C$-scheme  this can be seen by comparing  the central values of  (\ref{eq:del0FOCst}) and (\ref{eq:del0CIBC}), and  for $\overline{{\rm MS}}$ by comparing the central values of (\ref{eq:del0FOst}) and (\ref{eq:del0CIBst}). Note that the standard FOPT in the $C$-scheme  and the Borel-improved CIPT  in the $C$-scheme lead to results very close to the value obtained with the mathematical model \cite{BeJa} presented in the Appendix, which is $\delta^{(0)}=  0.2047 \pm 0.0029 \pm  0.0130$.

One can understand these results by a closer examination of the two expansions (a detailed discussion is given in \cite{CaFi2011}). The good predictions of the standard FOPT are actually due to some fortuitous cancellations between the contributions of large terms in the integral (\ref{eq:delta0}) along the circle. By the conformal mapping of the Borel plane, which improves the series  convergence, the large coefficients of the FO series near the timelike axis are no longer compensated to the same extent. Therefore, the  Borel-improved FO expansion is expected to give poorer results.  By contrast, in  the Borel-improved CI summation, the improvement of the series convergence is combined with the exact renormalization group summation of the running coupling along the circle $|s|=m_\tau^2$, ensuring a good convergence of the series (\ref{eq:hatDCI}) along the whole integration contour.  Therefore, FOPT appears to be the good choice for the standard expansions, while CIPT is the preferred choice for the Borel-improved expansions in both    $\overline{{\rm MS}}$ and $C$ schemes. With these options, the results of the FO and CI predictions for the $\tau$ decay width are compatible within errors.

\begin{figure}[htb]
\includegraphics[scale=0.30]{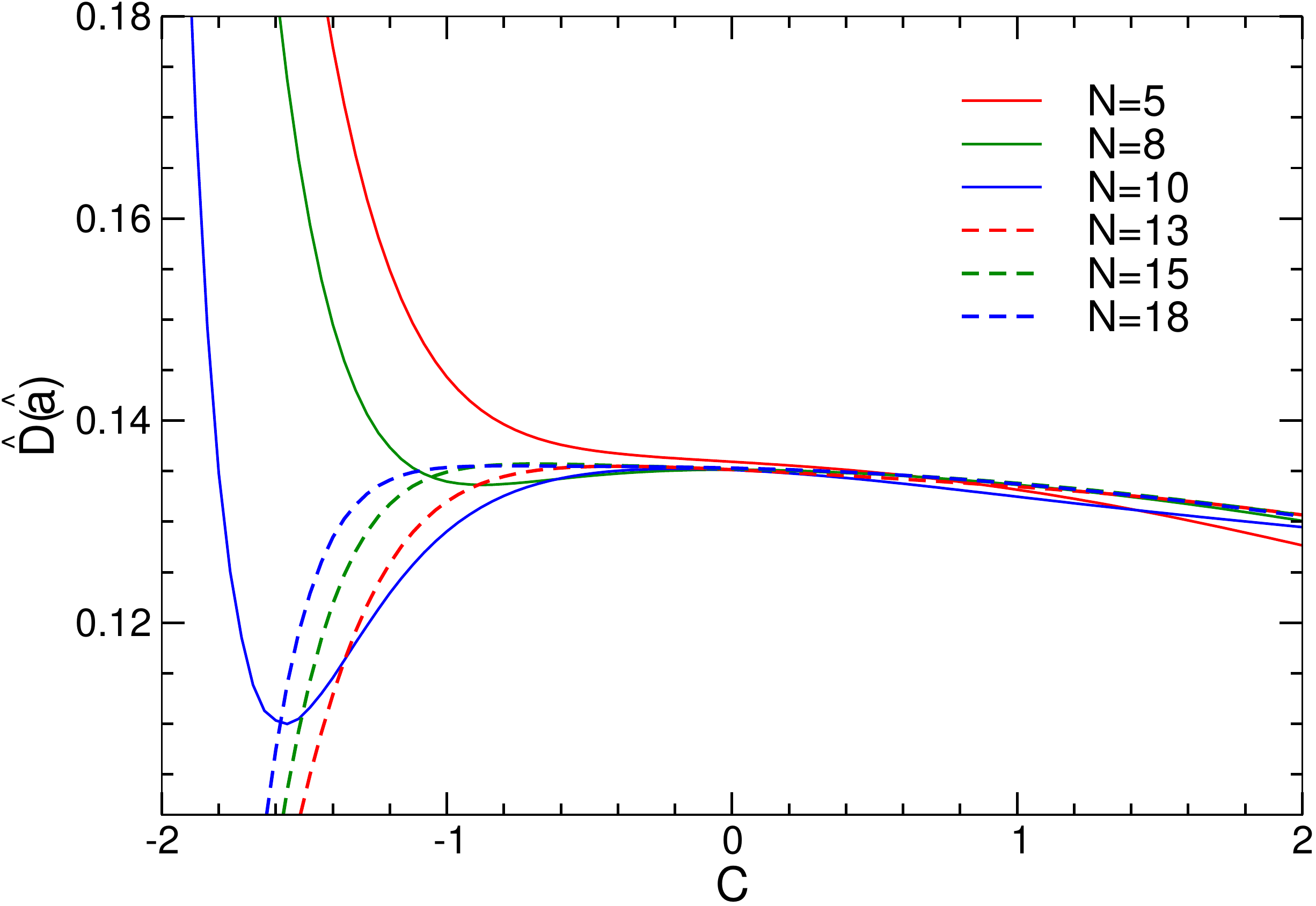}
\caption{Variation with $C$ of the Borel-improved expansions in the $C$-scheme of the Adler function $\wh D$ at $s=-m_\tau^2$, for various truncation orders $N$.\label{fig:4}}
\end{figure}

\subsection{High-order behavior}\label{sec:high}

 The high-order behavior of the standard QCD perturbation expansions in the $C$-scheme has been investigated in  \cite{Cscheme1, Cscheme2}, where a realistic renormalon-based model of the Adler function, proposed in \cite{BeJa},  was adopted for generating the higher-order perturbative coefficients. We use the same model (presented for completeness in the Appendix) for assessing the quality of the  Borel-improved expansions in the $C$-scheme.  We mention that the high-order behavior of the Borel-improved expansions in the $\overline{{\rm MS}}$ scheme has been discussed in  \cite{CaFi1998, CaFi2009, CaFi2011}. 

\begin{table*}[htb]
\caption{Adler function  $\wh D$ of the model \cite{BeJa} calculated at $s=-m_\tau^2$ with perturbative expansions in the $C$-scheme truncated at order $N$. Second column: standard expansion. Next three columns: expansions of Borel transform in powers of $u$ with first renormalons accounted for. Last four columns: expansions of the Borel transform in the optimal variable $w$, without and with renormalon softening.  Exact value of the model: $\wh D =0.1354 $.} \vspace{0.1cm}
\label{tab:1}
 \renewcommand{\tabcolsep}{0.5pc} 
\renewcommand{\arraystretch}{1.1} 
\begin{tabular}{l|c| ccc| cccc}\hline\hline
$N$ \,\,\,    & Eq. (\ref{eq:hatDC}) &Eq. (\ref{eq:Bir}) & Eq. (\ref{eq:Buv})  & Eq. (\ref{eq:Biruv})& Eq. (\ref{eq:Bw}) &   Eq. (\ref{eq:Bw2}) &  Eq. (\ref{eq:Bw3}) & Eq. (\ref{eq:Bw1})    \\\hline
3 & 0.1273   &  0.1329  & 0.1247  &0.1266 &0.1256 &  0.1401  &  0.1232 &  0.1280 \\                                   
4  & 0.1343  & 0.1379   & 0.1293 &0.1358 & 0.1441&    0.1372 &   0.1280 &             0.1360  \\
5  & 0.1343   & 0.1379 & 0.1331 &0.1358 & 0.1441 &    0.1372 &   0.1324 &     0.1360\\
6  & 0.1414   &  0.1268 &0.1359 &0.1371 &0.1435 &       0.1337 &   0.1361 &           0.1360 \\
7  & 0.1377  & 0.1560  &0.1388 & 0.1344&0.1439 &    0.1371 &    0.1391 &              0.1357 \\
8 &  0.1567   & 0.0949  &0.1421 & 0.1359&0.1363 &   0.1338 &   0.1412 &               0.1351\\ 
9  & 0.1283   & 0.1998 &0.1462 &0.1348 & 0.1253&     0.1352 &  0.1418    &           0.1350 \\
10 & 0.2279    & 0.0652  & 0.1515&0.1362 &0.1181&      0.1354 &   0.1405 &          0.1352\\ 
11 & -0.0195   &  0.1176 &0.1599 &0.1355 & 0.1177&     0.1347 &   0.1378 &              0.1351\\
12  & 0.8446   &  0.5601 & 0.1724 &0.1340 & 0.1192&     0.1355 &   0.1347 &            0.1351\\
13  & -2.004   & -1.607  & 0.1962& 0.1413& 0.1238&    0.1353 &    0.1326 &            0.1352\\
14  &8.982  & 5.581 & 0.2347 &0.1413 & 0.1300 &        0.1353 &    0.1321 &           0.1353  \\ 
15  &-34.61    &    -14.86 & 0.3288 &0.1706 &0.1305 & 0.1355 &     0.1332 &           0.1353\\
16  & 154.94  &   38.25& 0.4724 & 0.0594&0.1273 &   0.1353 &    0.1351 &              0.1353\\
17  & -711.57  & -90.63   &0.9929 &0.2902 &0.1247 &    0.1354 &  0.1367 &             0.1353 \\
18  & 3522.7    & 202.12 & 1.4979& -0.1629&0.1229&     0.1354 &     0.1372  &         0.1353 \\\hline
\hline \end{tabular}
\end{table*}

In Fig. \ref{fig:4} we present the variation with $C$ of the Borel-improved  expansions in the $C$-scheme of the Adler function $\wh D$  at the spacelike point $s=-m_\tau^2$,  for increasing orders of perturbation theory ($N$ denotes the number of terms kept in the expansion). In this calculation we  used the expansion (\ref{eq:Bw1}) of the Borel transform,  based on the optimal conformal mapping (\ref{eq:w}) and the softening factors expressed in the variable $w$.  The scale was fixed at $\mu=m_\tau$ and the perturbative coefficients of the model in  $\overline{{\rm MS}}$ scheme, given in Eq. (\ref{eq:cnBJ}), have been used as input.  

The curves shown in Fig. \ref{fig:4} exhibit a common region of stability with respect to $C$  and a remarkable convergence of the truncated expansions in this region.  The bigger variations which appear when $C$ is decreased towards the lower limit of the chosen range  are due to the fact that in this range the perturbative connection between the  QCD coupling in the $\overline{\rm MS}$ and $C$ schemes breaks down, so the use of perturbation theory is not legitimate.  

For a detailed numerical comparison, we present in Table \ref{tab:1} the values of the Adler function calculated at $s=-m_\tau^2$ with the scale  $\mu=m_\tau$, using several perturbative expansions in the $C$-scheme  discussed in Sec. \ref{sec:conf}. The aim was to assess the relative effect of the renormalon softening and the conformal mapping in improving the convergence. Thus, besides the standard perturbative expansion and the optimal expansion based on both conformal mapping and renormalon softening, we investigated expansions of $B(u)$ in powers of the Borel variable $u$, with the first UV renormalon removed, the    first IR renormalon removed and both UV and IR renormalons removed, as well as the expansion in the optimal conformal variable $w$ with no renormalon softening. The equations specifying these expansions are indicated for each column of Table   \ref{tab:1}.

An open problem in the analysis is the choice of the value of $C$ to be used in the expansions. Several values  have been considered in Refs. \cite{Cscheme1, Cscheme2} for the calculation of the Adler function and the quantity $\delta^{(0)}$ using the standard expansions in the $C$-scheme, which exhibit a divergent behavior. 

In the present work  we used, for each expansion, the value of $C$ determined in subsection \ref{sec:D} as the optimal $C$ for the corresponding expansion truncated at $N=5$, which means that the last term of the expansion with $N=5$ was set to zero. This explains why in Table  \ref{tab:1} the expansions with $N=4$ and $N=5$ coincide (except for columns 4 and 8, where no solution of this condition was found, and the value of $C$ minimizing the fifth term was adopted instead). 

 As shown in Fig. \ref{fig:4}, the Borel-improved expansions exhibit a common region of stability for all orders $N$, so the choice of a single $C$ in this region is reasonable at least for these expansions.
Of course, an optimal $C$ can be calculated for each truncation order $N$ (provided the condition has acceptable solutions), and an order-dependent $C$ can be used in practical applications. We recall that, while in the standard expansion the condition can be achieved only by the vanishing of the coefficient $\hat d_N$, in the expansions based on the Borel transform the expansion function $\hat{\cal W}_N$ itself can vanish. In our numerical study, it turned out that the  condition was achieved in most cases through the vanishing of the expansion function.

\begin{figure*}[htb]
\includegraphics[scale=0.30]{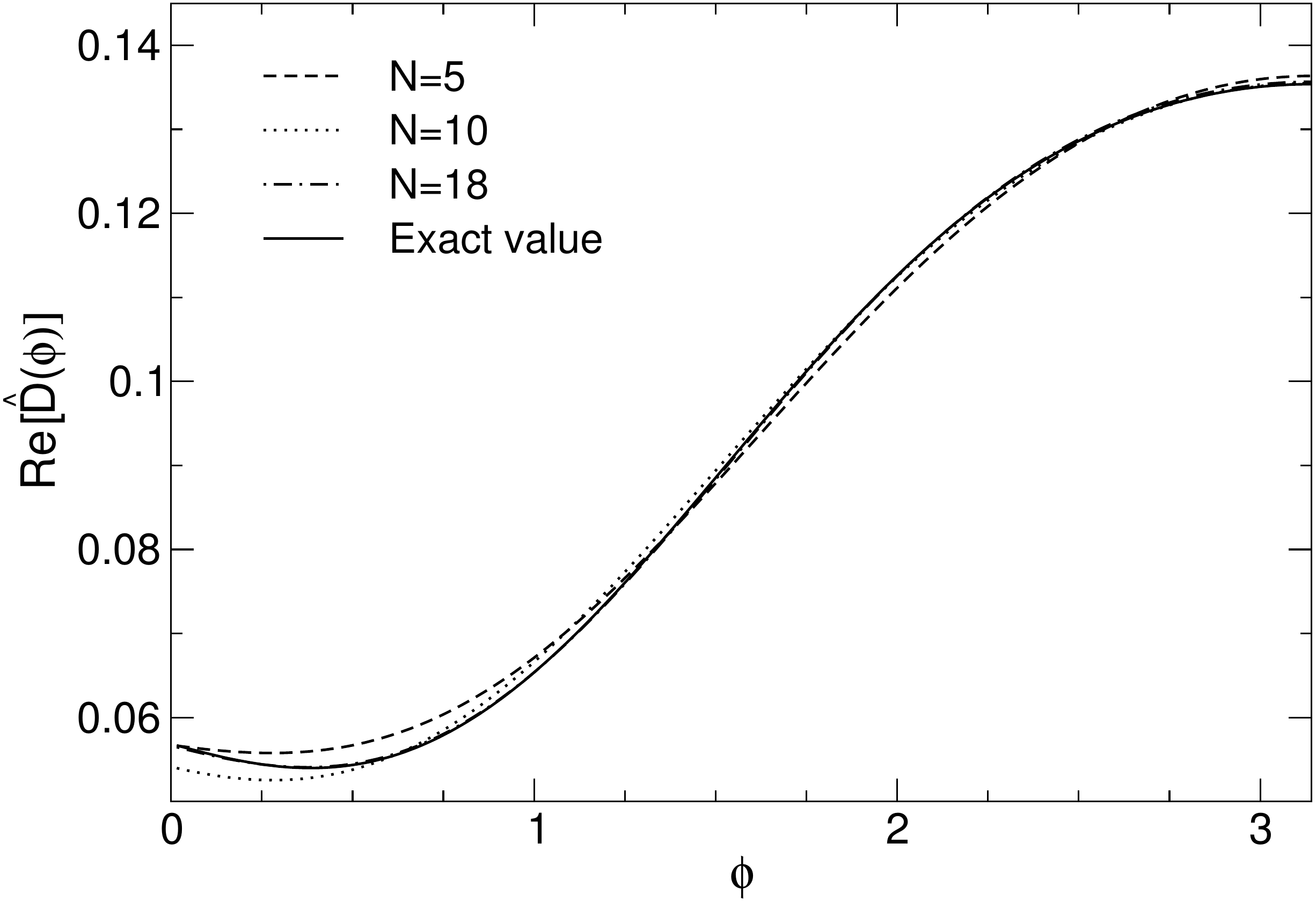}\hspace{0.7cm} \includegraphics[scale=0.30]{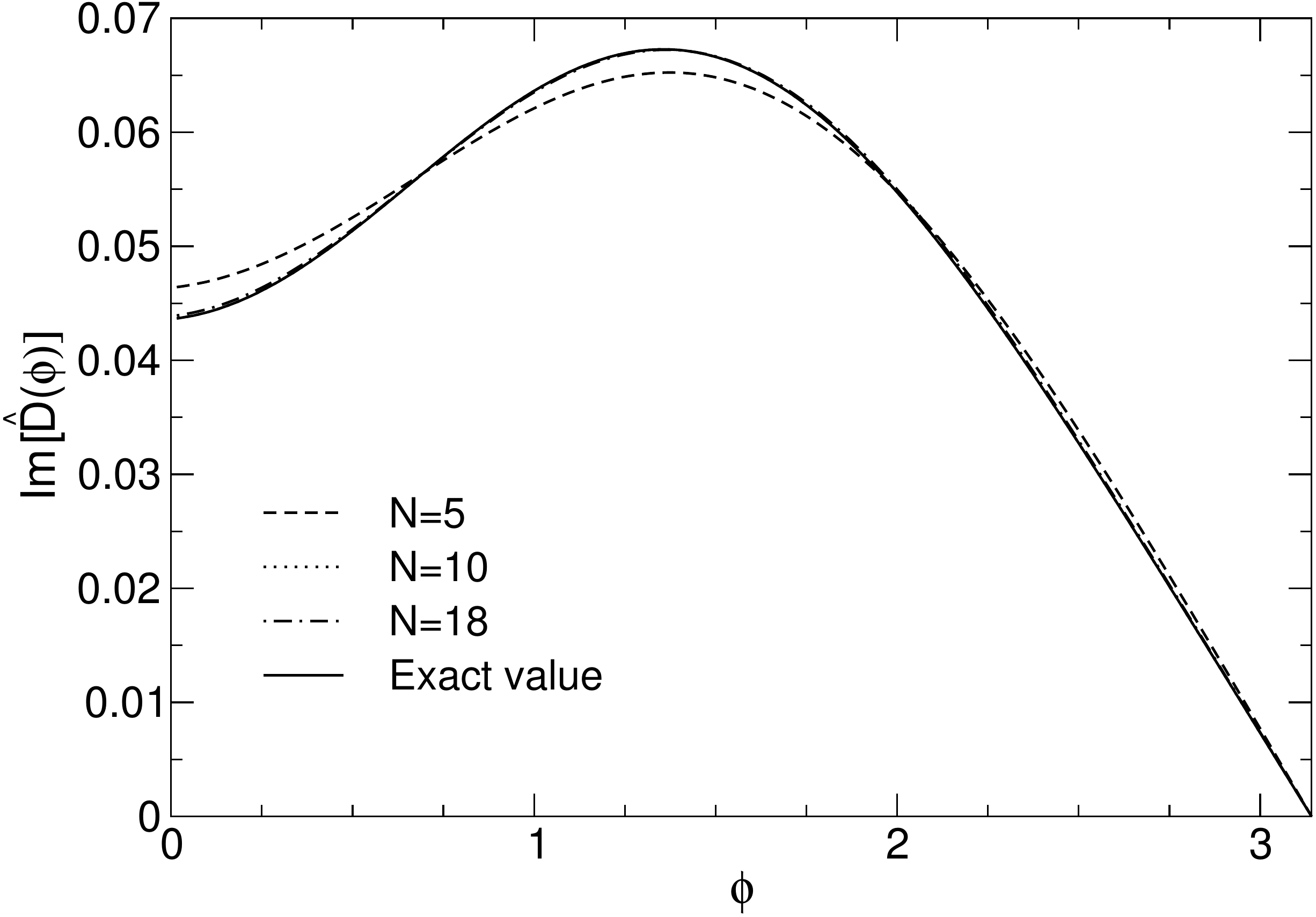}
\caption{Real part (left) and imaginary part (right) of the Adler function in the model \cite{BeJa} and its truncated Borel-improved expansions in the $C$-scheme along the circle $|s|=m_\tau^2 \exp(i \phi)$.\label{fig:5}}
\end{figure*}

From Table \ref{tab:1} one can remark the divergent behavior of the standard expansion in the $C$-scheme, given in column 2.  Column 3 shows that by softening the first IR renormalon at $u=2$,  the low perturbative orders are improved, but the series is badly divergent at large orders. This is expected actually, because the first UV renormalon, which is closest to the origin of the Borel plane, limits to $|u|<1$ the convergence region of the expansion (\ref{eq:Bir}).  From column 4 it is seen that the effect of softening the first UV renormalon is slightly weaker at low orders, but  the divergent character at larger orders is considerably tamed, while by softening both renormalons  good results are obtained up to relatively high orders. The expansion is however divergent, although this starts to be visible only at larger orders. Finally, from columns 6-9 it is seen that the effect of the conformal mapping is to ensure convergent expansions at high orders. The additional softening factors improve the description at low orders, the softening of the first IR renormalon having a more visible effect than that of the first UV renormalon. We note that by using  order-dependent optimal values of $C$, determined from the optimum condition at each order $N$,  the convergence of the expansions based on the conformal mapping  towards the exact result of the model is even more precise than shown in columns 6-9 of Table \ref{tab:1}.

 The large-order behavior of the expansions  of the $\tau$ hadronic width  has been also investigated,  in  the $\overline{\rm MS}$ scheme in \cite{BeJa, CaFi2009, CaFi2011}, and in the $C$-scheme in \cite{Cscheme1, Cscheme2}.  
As discussed in Sec. \ref{sec:Adler}, the quantity $\delta^{(0)}$,  which involves the values of the Adler function in the complex $s$ plane,  allows one to extract only indirect information about the perturbation expansion of the Adler function itself along the circle.

In order to assess in a straightforward way the quality of the expansions in the complex plane,  we compare in Fig. \ref{fig:5} the values of the Adler function calculated with the model presented in the Appendix for complex values  $s=m_\tau^2 \exp(i\phi)$, and its Borel-improved approximants in the  $C$-scheme given by (\ref{eq:DnewCIC}), with the running coupling $\ah_\mu$ calculated for $\mu^2=-s$ by integrating the renormalization-group equation in the $C$-scheme, as explained at the end of Sec. \ref{sec:confC}. We restricted $\phi$ to the range 
$\phi\in (0, \pi)$, the values on the semicircle in the lower half-plane being obtained by using the Schwarz reflection property  $\wh D(s^*)= \wh D^*(s)$.  The calculations have been done with the value $C=-0.05$, which is the optimal choice for the evaluation of the Adler function at $s=-m_\tau^2$ with $N=5$ terms in the expansion. One can remark the impressive convergence  along the whole circle $|s|=m_\tau^2$ of the Borel-improved expansions in the $C$-scheme, up to high perturbative orders. By inserting these expansions into the definition (\ref{eq:delta0}) of $\delta^{(0)}$, one obtains good predictions for this quantitity in CIPT, which, as discussed in the previous subsection, is the preferred choice of summation for the Borel-improved expansions.  As shown in \cite{CaFi2009, CaFi2011}, a similar behavior is obtained for the Borel-improved expansions in $\overline{\rm MS}$ scheme, while the standard expansions show big oscillations far from the true values.

\section{Discussion and conclusions}\label{sec:conc}
In the present paper we investigated the renormalization scheme variation of an improved perturbation expansion in QCD, with tamed behavior at large orders, defined by means of the optimal conformal mapping of the Borel plane. Detailed studies performed in previous works \cite{CaFi2009, CaFi_Manchester, CaFi2011, Abbas:2012fi, Abbas:2013}  demonstrated the good properties of these Borel-improved expansions in the $\overline{\rm MS}$ scheme. The analysis was extended now to a new renormalization scheme, denoted as $C$-scheme,  proposed in \cite{Cscheme} and investigated in the frame  of standard QCD expansions in  \cite{Cscheme, Cscheme1, Cscheme2}. Our purpose was not to advocate the advantage of a particular scheme, but to study the variation with the renormalization scheme of a quantity known to be renormalization-scheme invariant. We performed our study using as examples the perturbation expansions of the Adler function and of the $\tau$ hadronic width. 

For the expansions to order $\alpha_s^5$ and higher, we found  a range of stability of the results with respect to the variation of the  $C$ parameter defining the RS, shown in Figs. \ref{fig:1}-\ref{fig:4}.  In Ref. \cite{Cscheme}, the optimal value of $C$ was defined by requiring that the highest term in the expansion vanishes.
Moreover, in Ref. \cite{Cscheme}, an additional assumption was made, namely  the last nonzero term in the expansion was taken as the truncation error to be attached to the central prediction. We adopted in the present work the same prescriptions,  but found that they might be too rigid in some situations. Therefore, the values quoted as uncertainties in Sec. \ref{sec:res}  for the predictions of the Adler function and  the phenomenological parameter $\delta^{(0)}$ in the $C$-scheme must be taken with some caution. 

One problem is the danger of a distorted truncation uncertainty, due to accidentally large or small values of the last nonzero term. Another problem was revealed when studying the dependence of the results on the renormalization scale $\mu$: namely, if the optimal $C$ for the central value of $\mu$ is negative and large, it may not be suitable for  lower values of $\mu$, for which the coupling is too large (as seen from Fig. \ref{fig:0}), and the perturbation expansion breaks down. 
 We note also that the optimum  prescription  proposed in  \cite{Cscheme} cannot be applied to expansions at lower orders, $N=3$ and $N=4$. For the standard expansions in the $C$-scheme, no solution of the condition exists, as the three curves corresponding to Fig. \ref{fig:1} do not intersect each other. For the Borel-improved expansion in the  $C$-scheme, a common intersection point exists, but it is situated in a region of large negative $C$, where the variation with respect to $C$ is not stable.

These problems  suggest that a more elaborate definition of the optimal $C$ and of the theoretical error might be necessary. It would be reasonable, for instance,  to require the simultaneous fulfillment of several conditions: smallness of the last term in the expansion (instead of requiring an exact zero), stability with respect to the variation of $C$, and  stability with respect to the variation of the scale.  Such a study, necessary for  phenomenological  applications, in particular for a precise determination of $\alpha_s$ from hadronic $\tau$ decays, is beyond the scope of the present paper, whose main aim was to investigate the behavior of the Borel-improved expansions in the $C$-scheme. 

For the  parameter $\delta^{(0)}$, we found in the $C$-scheme a property already noticed in $\overline{{\rm MS}}$ scheme, namely the close results given by the CIPT Borel-improved  expansions and the FOPT standard expansions. Therefore, FOPT appears to be the good choice for the standard expansions, while CIPT is the preferred choice for the Borel-improved expansions in both    $\overline{{\rm MS}}$ and $C$ schemes.

The good properties of the  Borel-improved  expansions manifest themselves in an impressive way at large orders.  In our study,  for generating the higher perturbative coefficients  we used as in \cite{Cscheme, Cscheme1, Cscheme2} a theoretical model for the Adler function, proposed in \cite{BeJa}. The results shown in Figs. \ref{fig:4} and \ref{fig:5}  prove the remarkable convergence of the Borel-improved expansions in the  $C$-scheme of the Adler function evaluated on the Euclidian axis and in the complex $s$ plane. 

In order to assess the relative effects of the singularity-softening factors and of the conformal mapping of the Borel plane, we investigated also several other expansions, in which  only the singularity softening is incorporated, with no conformal mapping in the power expansion, or only the conformal mapping without singularity softening is used. The results given  in  Table \ref{tab:1} show that the proper treatment of lowest renormalons improves the low-orders but cannot cure the divergence at high orders, while the use of the optimal conformal mapping without softening factors ensures the convergence at high orders, but may give poorer results at low orders. Finally, the Borel-improved expansion based on  singularity-softening factors and the optimal conformal mapping of the Borel plane gives good results at low orders and converges towards the true values at large orders.

As discussed in Sec. \ref{sec:res}, an open problem for the expansions in the $C$-scheme is the proper choice of the parameter $C$ which defines a particular RS. We argued that an  order-dependent  optimal $C$ appears to be the best choice, and emphasized that in the Borel-improved expansion the optimum condition can be achieved also by imposing the vanishing of the last expansion function, a possibility that does not exist for the standard expansion.

We emphasize finally that  the model used  in the present study for generating higher-order perturbative coefficients was constructed in  \cite{BeJa} from a renormalon analysis in $\overline{{\rm MS}}$ scheme.  Moreover, the free parameters of the model are determined such as to generate  the lowest perturbative coefficients (\ref{eq:cn1}) known in  $\overline{{\rm MS}}$ scheme. However, we proved that the Borel-improved expansions converge to the exact result even if the expansions are defined in the $C$-scheme. This provides a nice illustration of  the renormalization-scheme independence of the QCD perturbation theory, once the large-order divergence is properly treated.

The results of the present analysis are a further argument in favor of the nonpower expansions based on the optimal conformal mapping of the Borel plane, which prove to be a useful tool for applications of perturbative QCD at intermediate energies.

\subsection*{Acknowledgments}
I thank D. Boito for useful discussions and suggestions on the manuscript. This work was supported by the Romanian Ministry of Research and Innovation, Contract PN 18090101/2018.
\appendix
\section{Model of the Adler function}
For testing the convergence of the various expansions, we considered the model proposed in \cite{BeJa}, which  expresses the Adler function  by means of the PV-regulated Laplace-Borel integral:  
\be\label{eq:pv}
\wh D(a_\mu)=\frac{1}{\beta_0}\,{\rm PV} \,\int\limits_0^\infty  e^{-\frac{u}{\beta_0 a_\mu}} \, B(u)\, {\rm d} u,
\ee
with a Borel transform $B(u)$ parametrized  in terms of a few UV and IR renormalons.
Specifically, in the model proposed in \cite{BeJa}, the function $B(u)$ is expressed as 
\be\label{eq:BBJ}
B(u)=B_1^{\rm UV}(u) +  B_2^{\rm IR}(u) + B_3^{\rm IR}(u) +d_0^{\rm PO} + d_1^{\rm PO} u,
\ee
 where
\be\label{eq:BIR}
B_p^{\rm IR}(u)= \frac{d_p^{\rm IR}}{(p-u)^{\gamma_p}}\,
\left[\, 1 + \tilde b_1 (p-u)  +\ldots \,\right],
\ee
\be\label{eq:BUV}
B_p^{\rm UV}(u)=\frac{d_p^{\rm UV}}{(p+u)^{\bar\gamma_p}}\,
\left[\, 1 + \bar b_1 (p+u)  +\ldots \,\right].
\ee

The free parameters of the models are the residues $d_1^{\rm UV}, d_2^{\rm IR}$ and  $d_3^{\rm IR}$ of the first renormalons and the coeficients $d_0^{\rm PO}, d_1^{\rm PO}$ of the polynomial in (\ref{eq:BBJ}), determined in  \cite{BeJa} as 
\bea 
&& d_1^{\rm UV}=-\,1.56\times 10^{-2},\,\,
d_2^{\rm IR}=3.16,\,\,
d_3^{\rm IR}=-13.5,\nn\\
&& d_0^{\rm PO}=0.781, \,\,
d_1^{\rm PO}=7.66\times 10^{-3},
\eea
by the requirement to reproduce the perturbative coefficients  $c_{n,1}$ in $\overline{{\rm MS}}$ scheme for $n\le 4$, given in (\ref{eq:cn1}), and the estimate $c_{5,1}=283$. 

Once the parameters are fixed, the model predicts all the higher order perturbative coefficients  $c_{n,1}$ for $n>5$. We give below the values of the coefficients used in the calculations presented in Sec. \ref{sec:res}:
singularity-softening factors and of the conformal mapping of the Borel plane
\bea\label{eq:cnBJ}
&& c_{6,1}=3275.45, \,\, c_{7,1}=18758.4, \,\, c_{8,1}=388446, \nn\\
 && c_{9,1}=919119, \,\,~~~~~~~~~~ c_{10,1}= 8.36\times 10^7,\nn\\\
&&c_{11,1}= -5.19\times 10^8, \,\, ~~ c_{12,1}=3.38\times 10^{10},  \nn\\ 
&& c_{13,1}= -6.04\times 10^{11},\,\, ~ c_{14,1}=2.34 \times 10^{13},\nn\\
&& c_{15,1}= -6.52 \times 10^{14},  \,\, ~ c_{16,1}=2.42 \times 10^{16}, \nn\\ 
&&c_{17,1}= -8.46\times 10^{17}, \,\, ~ c_{18,1}= 3.36 \times 10^{19}.
\eea
One can note the dramatic increase of the coefficients, which implies that the perturbation series of the Adler function in this model is divergent.

For $\alpha_s(m_\tau^2)=0.316\pm 0.010$,  the Adler function at $s=-m_\tau^2$ and $\mu=m_\tau$ given by this model has the value  \cite{Cscheme}: $\wh D(a_{m_\tau})=0.1354\pm 0.0127\pm 0.0058$, where the first error comes from renormalon ambiguity, evaluated using the prescription (\ref{eq:errorPV}) and the second from the uncertainty of the coupling.

\end{document}